\documentclass[fleqn]{article}
\usepackage{longtable}
\usepackage{graphicx}

\begin {document}
\begin{flushleft}
{\LARGE
{\bf Energy levels, radiative rates and electron impact excitation rates for transitions in Be-like Cl XIV, \linebreak[4]K XVI and Ge XXIX}
}\\

\vspace{1.5 cm}

{\bf {Kanti  M  ~Aggarwal and Francis  P  ~Keenan}}\\ 

\vspace*{1.0cm}

Astrophysics Research Centre, School of Mathematics and Physics, Queen's University Belfast, Belfast BT7 1NN, Northern Ireland, UK\\ 

\vspace*{0.5 cm} 

e-mail: K.Aggarwal@qub.ac.uk \\

\vspace*{1.50cm}

Received  31 July 2014\\
Accepted for publication 26 August 2014 \\
Published xx  Month 2014 \\
Online at stacks.iop.org/PhysScr/vol/number \\

\vspace*{1.5cm}

PACS Ref: 32.70 Cs, 34.80 Dp, 95.30 Ky

\vspace*{1.0 cm}

\hrule

\vspace{0.5 cm}
{\Large {\bf S}} This article has associated online supplementary data files \\
Tables 4--6 and 13--15 are available only in the electronic version at stacks.iop.org/PhysScr/vol/number

\end{flushleft}

\clearpage


\begin{abstract}

Results for energy levels, radiative rates and electron impact excitation (effective)  collision strengths  for transitions in Be-like Cl XIV, K XVI and Ge XXIX are reported.  For the calculations of energy levels and radiative rates the General-purpose Relativistic Atomic Structure Package ({\sc grasp}) is adopted, while for determining the collision strengths and subsequently the excitation rates, the Dirac Atomic R-matrix Code ({\sc darc}) is used.  Oscillator strengths, radiative rates and line strengths are listed for all E1, E2, M1 and M2 transitions among the lowest 98 levels of the $n \le$ 4 configurations. Furthermore,  lifetimes are provided for all  levels and comparisons made with available theoretical and experimental results.  Resonances in the collision strengths are resolved in a fine energy mesh and  averaged over a Maxwellian velocity distribution to obtain the effective collision strengths. Results  obtained are listed over a wide temperature range up to 10$^{7.8}$ K, depending on the ion.  

\end{abstract}

\clearpage

\section{Introduction}

Coronal lines of Be-like ions are useful for plasma diagnostics. In particular, the intensity ratio of the 2s$^2$ $^1$S$_0$ -- 2s2p $^3$P$^o_1$ and 2s2p $^1$P$^o_1$ -- 2p$^2$ $^1$D$_2$ lines is temperature sensitive, as demonstrated by Landi {\em et al} \cite{el} for a range of ions with 6 $\le$ Z $\le$ 28. Similarly, the intensity ratio of the  2s2p $^3$P$^o_2$ -- 2p$^2$ $^3$P$_2$ and 2s$^2$ $^1$S$_0$ -- 2s2p $^1$P$^o_1$ lines is density sensitive \cite{fpk2}. While these lines are in the UV range, the magnetic dipole (2s2p) $^3$P$^o_1$ --  $^3$P$^o_2$ line of 17 $\le$ Z $\le$ 20 lies in the visible  \cite{pdb}.  In particular, lines of K XVI have been  detected in solar spectra from the SERTS (solar EUV rocket telescope and spectrograph) rocket \cite{tn} as well as {\em Hinode} satellite \cite{del}. Apart from astrophysical applications, Be-like ions are of interest in the modelling of fusion plasmas  \cite{fs}. Several lines of Cl XIV and K XVI have been measured in laboratory plasmas by Huang {\em et al} \cite{lkh}. Similarly, the 2s$^2$ $^1$S$_0$ -- 2s2p $^1$P$^o_1$ line of Ge XXIX at 92.8 $\rm \AA$ has been detected in the PLT (Princeton Large Torus) tokamak  plasma by Stratton {\em et al} \cite{plt}. The importance of these ions has further increased with the developing ITER project. The ions which have been identified \cite{sjr} for a greater interest are Al X, Cl XIV, K XVI, Ti XIX and Ge XXIX. Therefore, we have recently reported atomic data (i.e.  energy levels, radiative rates and electron impact excitation rates) for Al X \cite{alx} and Ti XIX \cite{tixix}, and in this paper report similar results for Cl XIV, K XVI and Ge XXIX.

The experimental  energy levels for these ions have been compiled and critically evaluated by the  NIST (National Institute of Standards and Technology) team \cite{nist}, and are available at the  website \\{\tt http://www.nist.gov/pml/data/asd.cfm}. Calculations for radiative rates (A- values) have been performed by many workers \cite{zs92} -- \cite{uis3}, but are restricted to the $n \le$ 3 levels. However, realising the importance of Be-like ions, many measurements have been made for their lifetimes, such as by  \cite{ishi}--\cite{et}. These are helpful in assessing the accuracy of calculated A- values, as we will discuss in section 4. On the other hand, calculations for collision strengths ($\Omega$) are very limited. For example, \cite{zs92} and \cite{zs84} calculated $\Omega$ for a wide range of Be-like ions with 8 $\le$ Z $\le$ 92, but for a limited range of transitions within the $n$ = 3 levels. Furthermore, unfortunately they did not report results for  ions of Cl, K or Ge. However, Keenan \cite{fpk1}  has provided analytical expressions, for transitions among the lowest 10 levels of the $n$=2 configurations,  which allow the determination  of effective collision strengths ($\Upsilon$) for Cl XIV and K XVI,  based on  interpolation of $R$-matrix calculations for three Be-like ions, namely Ne VII, Si XI and Ca XVII. These results have been used in plasma modelling  \cite{fpk2} and are  stored in the CHIANTI \cite{chianti} database at ${\tt {\verb+http://www.chiantidatabase.org+}}$. Since these data are not based on direct calculations for these ions, the available results can at best be described as preliminary, and hence there is  scope  for improvement as well as extension, as was the case with Al X \cite{alx}. Finally, for Ge XXIX the only calculations available in the literature for $\Omega$ are by \cite{akb},  based on the {\em distorted-wave} (DW) method,  which do not include the contribution of resonances. Also data for  $\Omega$ are reported at only a single energy,  and therefore the subsequent values of $\Upsilon$ may be significantly underestimated, because for Be-like ions the resonance structure is very  dense, as demonstrated in our work on Ti XIX \cite{tixix}. Therefore,  here we report atomic data for energy levels, A- values, $\Omega$ and $\Upsilon$ for transitions among the lowest 98 levels of the $n \le$ 4 configurations of Cl XIV,  K XVI  and Ge XXIX. 

As in most of our previous work, including on Be-like Al X \cite{alx} and Ti XIX \cite{tixix}, the fully relativistic {\sc grasp} (general-purpose relativistic atomic structure  package) code has been used to calculate energy levels and A- values. Several versions of this code are available in the literature, but the one adopted here has been revised by  one of the authors (Dr P H Norrington) and is freely available at the website  {\tt http://web.am.qub.ac.uk/DARC/}). This code is based on the $jj$ coupling scheme,  and apart from the one-body relativistic terms also includes corrections arising from the Breit interaction and QED effects (vacuum polarization and Lamb shift). Similarly, as in earlier research we have used the option of {\em extended average level} (EAL),  in which a weighted (proportional to 2$j$+1) trace of the Hamiltonian matrix is minimized. This produces a compromise set of orbitals describing closely lying states with  moderate accuracy. For our calculations of $\Omega$, we have adopted the {\em Dirac Atomic $R$-matrix Code} ({\sc darc}) of P H Norrington and I P Grant ({\tt http://web.am.qub.ac.uk/DARC/}). 

\section{Energy levels}

The 17 configurations of Be-like ions, namely (1s$^2$) 2$\ell$2$\ell'$, 2$\ell$3$\ell$ and 2$\ell$4$\ell$,  generate 98 levels. These are  listed in Tables 1--3 for Cl XIV, K XVI and Ge XXIX, respectively. Also included in these tables are  our level energies calculated from {\sc grasp},  {\em without} (GRASP1) and {\em with} (GRASP2) the contributions of the  Breit and QED effects, as well as the values compiled by the NIST team \cite{nist}.   However, NIST energies are restricted to the $n \le$ 3 levels and even then are not available for all levels, particularly  in the case of  Ge XXIX.  As expected, the contributions of Breit and QED effects are negligible ($\le$ 0.04 Ryd) for Cl XIV and K XVI, but are slightly  significant (up to 0.2 Ryd) for the comparatively heavy ion Ge XXIX -- see for example, the $n$ = 4 levels in Table 3. For a majority of levels, the energies have become lower with the inclusion of Breit and QED effects, but have increased for a few, particularly for  2s2p $^3$P$^o_0$. 

Our  energies obtained with the inclusion of the Breit and QED effects (GRASP2) are generally in good agreement (within 0.08 Ryd) with the NIST listings for most of the common levels. However, there are some exceptions. For example, for levels 5, 9 and 10 of Cl XIV, the NIST energies are lower by $\sim$ 0.1 Ryd and higher by the same amount for level 27 (2p3p $^3$D$_2$). Similar differences are noted for the same levels of K XVI, as found earlier for Al X \cite{alx} and Ti XIX \cite{tixix}. Since NIST energies are not available for a majority of levels, the comparisons shown in Tables 1--3 are not very helpful for assessing the accuracy of our calculations. Therefore, also listed in these tables are the theoretical energies of Gu \cite{gu} for the lowest 10 levels and Safronova {\em et al} \cite{uis4} for the higher ones,  obtained from  relativistic  {\em many-body perturbation theory} (MBPT). In general, agreement between the GRASP and MBPT energies is highly satisfactory (within 0.1 Ryd) for a majority of levels and for all three ions. However, the MBPT energies are (mostly) closer to the NIST listings, although differences between the two are up to 0.1 Ryd (see for example, level 27 of Cl XIV in Table 1), i.e. the discrepancies are the same as with the GRASP calculations. We discuss these further. 

Unpublished results of C Froese Fischer (2009), based on MCHF (multi-configuration Hartree-Fock) calculations are also available for 35 and 24 levels of Cl XIV and K XVI, respectively, on the website\\ {\tt http://nlte.nist.gov/MCHF/view.html}. Therefore, in table 1  we have included her results for Cl XIV for comparison. These energies are closer to the NIST listings and discrepancies, if any, are within 0.01 Ryd. Similarly, there is a good agreement between the MCHF and MBPT energies, and this clearly indicates that there is scope for improvement in our calculated energies with the {\sc grasp} code. This is further confirmed by the large-scale calculations of Cheng {\em et al} \cite{ccj}, based on relativistic configuration-interaction (RCI) method, who included over 200,000 configurations to determine energies of the lowest 4 levels of Be-like ions with Z $\le$ 92. As with the MBPT and MCHF calculations, their energies are in complete agreement, for all three ions under discussion here, with the experimental compilations of NIST. Finally, J\"{o}nsson {\em et al} \cite{jgg} have demonstrated that the {\em convergence} in energy levels (and A- values), and a closer agreement with the experimental results, can be achieved for a range of ions, including Be-like, but with a very large expansion  of configuration state functions (CSFs). For example, they included over 800,000 and 1,000,000 CSFs for the even and odd states of O IV to determine the $n \le$ 3 energy levels within 0.25\% of the measurements. However, it must be stressed that such large calculations, although desirable, are not practically possible (particularly with the computational resources available with us) for a large number of levels, and especially when measurements are not available to compare with and to make accuracy assessment. Furthermore, the main focus in our work is on collisional data (discussed in sections 5 and 6) and for which a large set of CSFs cannot be included.  

Since the comparisons of energies discussed above are restricted to the $n \le$ 3 levels and no prior results are available for the $n$ = 4 levels,   we have performed parallel calculations with the {\em Flexible Atomic Code} ({\sc fac}) of Gu \cite{fac}, available from the website {\tt {\verb+http://sprg.ssl.berkeley.edu/~mfgu/fac/+}}.  {\sc fac}, like {\sc grasp},  is also a fully relativistic code which provides a variety of atomic parameters, including energy levels. As shown in some of our  earlier papers (see for example, \cite{tixix} and \cite{fe15}),  results for energy levels and radiative rates obtained from {\sc fac} are comparable to those from {\sc grasp}.   Hence calculations from {\sc fac} are helpful in assessing the accuracy of our energy levels, as well as radiative rates (which will be discussed in the next section).  Therefore, listed in Tables 1--3 are results obtained from the {\sc fac} code (FAC1), which include the same CI (configuration interaction) as in the {\sc grasp} calculations.  

Our FAC1 level energies  agree with our GRASP2 data within 0.05 Ryd for all levels and all ions. More importantly, the orderings are also the same.  For some ions, a larger expansion of CI improves the accuracy of energy levels. To test this we have performed yet another calculation (FAC2), which includes a further 68 levels of the  2$\ell$5$\ell$ configurations. The FAC2 results are also included  in Tables 1--3, but differences with the $n$ = 4 calculations (FAC1) are negligible and the orderings  of the levels are also the same. However, this (FAC2) expansion is insignificant in comparison to the large calculations \cite{ccj}--\cite{jgg} discussed above. Nevertheless, based on the comparisons discussed above we are confident that our energy levels listed in Tables 1--3 for Cl XIV, K XVI and Ge XXIX are accurate to better than 0.1 Ryd, or equivalently $\sim$ 2\%. The accuracy of energy levels can be improved with the inclusions of {\em core-valence} and {\em core-core} correlations, apart from the {\em valence-valence} interactions included in our work. Similarly, orbitals of higher $n$ will be helpful in improving the accuracy. Finally, we may state that all sets of energies, experimental and theoretical,  agree within 0.1 Ryd and there is no discrepancy in their orderings, but scope remains for improving the accuracy of energy levels listed in tables 1--3.

 \section{Radiative rates}

We have calculated  A- values  for four types of transition among the 98 levels, namely electric dipole (E1), electric quadrupole (E2), magnetic dipole (M1) and  magnetic quadrupole (M2). Generally, A- values for E1 transitions dominate  and hence are the most important. However, occasionally other types of transitions are also significant, such as  (2s2p) $^3$P$^o_1$ --  $^3$P$^o_2$ (M1), as mentioned in section 1.  Therefore, for a complete  plasma model, the inclusion of  A- values for all types of transitions is generally preferable. In Tables 4--6 we list transition energies/wavelengths ($\lambda$, in $\rm \AA$), radiative rates (A$_{ji}$, in s$^{-1}$), oscillator strengths (f$_{ij}$, dimensionless), and line strengths (S, in a.u.), in length  form only, for all 1468 electric dipole (E1) transitions among the 98 levels of Cl XIV, K XVI and Ge XXIX. The {\em indices} used  to represent the lower and upper levels of a transition have already been defined in Tables 1--3. Similarly, there are 1754 electric quadrupole (E2), 1424  magnetic dipole (M1) and 1792 magnetic quadrupole (M2) transitions among the 98 levels. However, for these transitions only the A- values are listed in these tables, because the corresponding results for f- or S- values can be easily obtained using Eqs. (1--5) of \cite{tixix}. Furthermore, the absorption oscillator strength ($f_{ij}$) and radiative rate A$_{ji}$ (in s$^{-1}$) for all types of  transitions  are related by the following expression:

\begin{equation}
f_{ij} = \frac{mc}{8{\pi}^2{e^2}}{\lambda^2_{ji}} \frac{{\omega}_j}{{\omega}_i} A_{ji}
 = 1.49 \times 10^{-16} \lambda^2_{ji}  \frac{{\omega}_j}{{\omega}_i} A_{ji}
\end{equation}
where $m$ and $e$ are the electron mass and charge, respectively, $c$ is the velocity of light,  $\lambda_{ji}$ is the transition energy/wavelength in $\rm \AA$, and $\omega_i$
and $\omega_j$ are the statistical weights of the lower ($i$) and upper ($j$) levels, respectively. 

Very limited data are available in the literature with which to compare our A- or f- values. Cheng {\em et al} \cite{ccj} have reported A- values for 3 transitions of Be-like ions, namely 1--3 E1, 1--4 M2 and 1--5 E1, and their results agree within 10\% with our calculations for all three ions. However, as for energy levels, unpublished results of C Froese Fischer (2009) are available on the website {\tt http://nlte.nist.gov/MCHF/view.html} for 175 E1 transitions among the $n \le$ 3 levels of Cl XIV and K XVI. These calculations are based on her {\em multi-configuration Hartree-Fock} (MCHF) theory. Therefore, in Table 7 we compare our f- values with her data for both ions, but only for transitions from the lowest 5 levels. Also included in this table are the ratio (R) of the velocity and length forms (i.e. the Coulomb and Babushkin gauges in the relativistic terminology), because they give an indication of the accuracy of the calculations. As for the energy levels, agreement between the two sets of f- values is highly satisfactory and the discrepancies, if any, are under 20\% for most transitions. There are slightly larger discrepancies (up to 50\%)  for some weak transitions, such as 3--12 (f $\sim$ 10$^{-6}$) of Cl XIV. Similarly, for most transitions with significant  f- values ($>$ 0.01) R is within 20\% of unity, although for a few weaker transitions (such as 5--6, f $\sim$ 10$^{-5}$) the two forms of f- value differ considerably. Weak(er) transitions are very sensitive  to different levels of CI (and methods), because of the cancellation or addition of the mixing coefficients. Such large discrepancies are therefore common for almost all ions, with examples in both  Al X \cite{alx} and Ti XIX \cite{tixix}.

The comparison of f- values discussed above is only for  about 12\% of the transitions, because of a paucity of other similar results. Therefore, as for energy levels, we have performed another calculation with the {\sc fac} code of Gu \cite{fac}.  For all {\em three}  ions the f- values from {\sc grasp} and {\sc fac} agree to within 20\% for all strong transitions, as was also noted for other Be-like ions, see for example Table 3 for Al X \cite{alx} and Ti XIX \cite{tixix}. Similarly, the effect of additional CI with the $n$ = 5 configurations (166 levels in total) is negligible for most transitions of all ions. Therefore, based on the comparisons discussed above, and earlier ones for other Be-like ions, we may confidently state that the accuracy of our radiative data listed in Tables 4--6 is better than 20\% for a majority of the (strong) transitions.

As for the E1 transitions, very limited data are available for comparison  for other types of transition. In Table 8 we compare A- values for Cl XIV and K XVI for one M2 (1--4) and 11 M1 transitions with the MBPT calculations of  Safronova \cite{uis5} and  Safronova {\em et al} \cite{uis3}. Considering that most such transitions are very weak (f $\sim$ 10$^{-6}$ or less), the agreement between the two sets of calculations is highly satisfactory, i.e. normally within 20\%. Only for two transitions (7--9 and 8--9) are the discrepancies for both ions  $\sim$50\%. Overall, our listed A- values in Tables 4--6 for all types of transition are estimated to be accurate and reliable.

\section{Lifetimes}

The lifetime $\tau$ for a level $j$ is defined as follows:

\begin{equation}  {\tau}_j = \frac{1}{{\sum_{i}^{}} A_{ji}}.  
\end{equation} 
 Since several measurements for a few levels of Cl XIV and K XVI are available in the literature,  in Tables 1--3 we have also listed our calculated lifetimes. Contributions from all four types of transitions, i.e. E1, E2, M1 and M2 are included for greater accuracy. In Table 9 we compare our calculated $\tau$ with  measurements for the lowest 10 levels of Cl XIV. For all levels there are no discrepancies between theory and experiment, which confirms, yet again, the accuracy of our calculated A- values. Also included in this table are the MBPT  results of Safronova {\em et al} \cite{uis2} (and of Andersson {\em et al} \cite{azh} for level 5) which appear to be overestimated for the 2p$^2$ $^3$P$_{0,1,2}$ levels by $\sim$50\%,  compared to measurements as well as with the present calculations. Finally, Tr\"{a}bert {\em et al} \cite{et} measured $\tau$ for the 2s2p $^3$P$^o_{2}$ level of K XVI to be 7.6$\pm$0.5 ms, and this compares very well with our result of 7.44 ms.
 
 \section{Collision strengths}

The collision strength ($\Omega$) and  collision cross section ($\sigma_{ij}$, $\pi{a_0^2}$)  are related by the following:

\begin{equation}
\Omega_{ij}(E) = {k^2_i}\omega_i\sigma_{ij}(E)
\end{equation}
where ${k^2_i}$ is the incident energy of the electron and $\omega_i$ is the statistical weight of the initial state. Results for collisional data are presented here
in the form of $\Omega$ as it is a symmetric and dimensionless quantity.

The {\em Dirac atomic $R$-matrix code} ({\sc darc}), employed for the computation of collision strengths $\Omega$,  includes the relativistic effects in a
systematic way, in both the target description and the scattering model. It is based on the $jj$ coupling scheme, and uses the  Dirac-Coulomb Hamiltonian in the $R$-matrix
approach. The $R$-matrix radii adopted for Cl XIV, K XVI and Ge XXIX  are  4.80, 4.40 and 2.40 au, respectively. For the expansion of the wavefunction,  55  continuum orbitals have been included for each channel angular momentum, which  allow us to compute $\Omega$ up to  energies  of 660, 780 and 2500 Ryd for Cl XIV, K XVI and Ge XXIX, respectively. These energy ranges are  sufficient to calculate values of effective collision strength $\Upsilon$ (see section 6)  up to T$_e$ = 10$^{7.8}$ K, well in excess of  the temperature of maximum abundance in ionisation equilibrium for the ions under consideration  \cite{pb}.  The maximum number of channels for a partial wave is 428, and the corresponding size of the Hamiltonian matrix is 23 579. To obtain convergence of  $\Omega$, we have included all partial waves with angular momentum $J \le$ 40.5. This wide range of partial waves is sufficient for convergence of $\Omega$ for most transitions (particularly forbidden and inter-combination) and  energies -- see Figs. 1--3 of \cite{tixix}. Furthermore,  to account for higher neglected partial waves, we have included a top-up, based on the Coulomb-Bethe approximation \cite{ab} for allowed transitions and geometric series for others. 

In Tables 10--12 we list our values of $\Omega$ for resonance transitions of Cl XIV, K XVI and Ge XXIX at energies {\em above} thresholds. The  indices used  to represent the levels of a transition have already been defined in Tables 1--3. Unfortunately, no similar data are available for comparison as already noted in section 1. As for energy levels and A- values, data for $\Omega$ can also be calculated with the {\sc fac} code. However, these data are not very useful for comparisons because there are often anomalies in the calculated $\Omega$, as may be seen in Fig. 6  of \cite{mgxi}--\cite{caxix}. Nevertheless, the qualitative agreement for most transitions in all three ions is similar to that already discussed for Al X \cite{alx} and Ti XIX \cite{tixix}. 

\section{Effective collision strengths}

Excitation rates, as well as energy levels and A- values, are required for plasma modelling, and are determined from the collision strengths ($\Omega$). Since the threshold energy region is dominated by numerous closed-channel (Feshbach) resonances, as shown in Figs. 6--11 of \cite{tixix}, values of $\Omega$ need to be calculated in a fine energy mesh  so that their contribution can be included, which is often significant (if not dominant) particularly for  forbidden and inter-combination transitions. Furthermore, for most (astrophysical and fusion) plasmas electrons have a wide distribution of velocities, and therefore values of $\Omega$ are  averaged over a
{\em Maxwellian} distribution as follows:

\begin{equation}
\Upsilon(T_e) = \int_{0}^{\infty} {\Omega}(E) {\rm exp}(-E_j/kT_e) d(E_j/{kT_e}),
\end{equation}
where $k$ is Boltzmann's constant, T$_e$  electron temperature in K, and E$_j$  the electron energy with respect to the final (excited) state. Once the value of $\Upsilon$ is
known the corresponding results for the excitation q(i,j) and de-excitation q(j,i) rates can be easily obtained from the following equations:

\begin{equation}
q(i,j) = \frac{8.63 \times 10^{-6}}{{\omega_i}{T_e^{1/2}}} \Upsilon {\rm exp}(-E_{ij}/{kT_e}) \hspace*{1.0 cm}{\rm cm^3s^{-1}}
\end{equation}
and
\begin{equation}
q(j,i) = \frac{8.63 \times 10^{-6}}{{\omega_j}{T_e^{1/2}}} \Upsilon \hspace*{1.0 cm}{\rm cm^3 s^{-1}},
\end{equation}
where $\omega_i$ and $\omega_j$ are the statistical weights of the initial ($i$) and final ($j$) states, respectively, and E$_{ij}$ is the transition energy.  Values of $\Omega$ need to be determined over a wide energy range (above thresholds)  to obtain convergence of the integral in Eq. (4), as clearly demonstrated in Fig. 7 of Aggarwal and Keenan \cite{ni11a}. 

To resolve resonances, we have performed our calculations of $\Omega$ at over $\sim$ 90 660 energies in the thresholds region, depending on the ion. Close to thresholds  the energy mesh is 0.001 Ryd and is 0.002 Ryd elsewhere, particularly when the energy gap between two thresholds is rather wide -- see for example levels 10 and 11 in Tables 1--3. Hence care has been taken to include as many resonances as possible, and with as fine a resolution as  computationally feasible. The density and importance of resonances can be appreciated from Figs. 6--11 of \cite{tixix}. For the present ions  we observe similar resonances, but for brevity we show these in Figs. 1a--3a for only three transitions of Ge XXIX, namely  1--2 (2s$^2$ $^1$S$_0$ -- 2s2p $^3$P$^o_0$),  1--4 (2s$^2$ $^1$S$_0$ -- 2s2p $^3$P$^o_2$) and  2--3 (2s2p $^3$P$^o_0$ -- 2s2p $^3$P$^o_1$). Because of the large vertical scales of the figures the importance of the resonances appears to be insignificant, except for the 1--4 transition. Therefore, to appreciate their importance we make similar plots in Figs. 1b--3b, but with limited vertical scales.  Although Ge XXIX is comparatively a heavy ion,  resonances are as dense, for all three transitions and many more,  over the entire thresholds energy range as for lighter ions, and therefore make a significant contribution to  $\Upsilon$  over a wide range of  temperature. 

Our calculated values of $\Upsilon$ are listed in Tables 13--15 over a wide temperature range up to 10$^{7.8}$ K, suitable for applications to a variety of plasmas. As stated in section 1, the only  available data for comparison are those of \cite{fpk1}, which are obtained from the interpolation of $R$-matrix calculations for Ne VII, Si XI and Ca XVII.  Therefore, in Tables 16 and 17 we  compare our values of $\Upsilon$ with those of \cite{fpk1} for transitions among the lowest 10 levels, and at three temperatures  of Cl XIV (log T$_e$ = 6.3, 6.5 and 6.7 K) and K XVI  (log T$_e$ = 6.5, 6.7 and 6.9 K).  Also included for easy guidance in these tables are the ratio of our and the $\Upsilon$ values of \cite{fpk1} at all three temperatures. As expected, agreement between the two sets of results for (most) allowed transitions is highly satisfactory for both ions -- see for example, 1--5, 2--7 and 3--6 in Tables 16--17. However, for forbidden transitions, such as 1--6/7/8 (2s$^2$ $^1$S$_0$ -- 2p$^2$ $^3$P$_{0,1,2}$) and 2--8/9/10 (2s2p $^3$P$^o_0$ -- (2p$^2$) $^3$P$_2$, $^1$D$_2$, $^1$S$_0$), the $\Upsilon$ of \cite{fpk1} are underestimated by up to a factor of 4. A similar effect of resonances (and/or discrepancy between two sets of $\Upsilon$) was observed for transitions of Al X \cite{alx}, and is fully expected considering the resonances observed in Figs. 1--3.

\section{Conclusions}

In this paper we have presented results for energy levels and  radiative rates for four types of transitions (E1, E2, M1 and M2) among the lowest 98 levels of the $n \le$ 4 configurations for three Be-like ions, namely  Cl XIV, K XVI and Ge XXIX. These ions are of interest for both astrophysical and fusion plasmas. Additionally, lifetimes of all the calculated levels have been reported, and these compare well with measurements.  Therefore, based on this and a variety of other comparisons among various calculations, including analogous ones with the {\sc fac} code,  our GRASP results for radiative rates, oscillator strengths, line strengths and  lifetimes  are assessed to be accurate to better than 20\% for a majority of the strong transitions (levels). Similar comparisons for collision strengths are not possible due to a paucity of data in the literature. However, based on our experience with similar calculations for other Be-like ions, namely Al X \cite{alx} and Ti XIX \cite{tixix}, the accuracy of our values of  $\Omega$s is $\sim$20\% for most transitions.  For  calculations of  effective collision strength ($\Upsilon$), resonances in the thresholds energy region are resolved in a fine energy mesh and are noted to be dominant for many transitions. Their contribution to the calculation of $\Upsilon$ is hence significant for all three ions of present interest.  Furthermore,  we have considered a large range of partial waves to achieve convergence of $\Omega$ at most energies and have adopted  a wide energy range to  calculate  values of $\Upsilon$ up to T$_e$ = 10$^{7.8}$ K.  Therefore,  we estimate the accuracy of our results for  $\Upsilon$  to be better than 20\% for most transitions.  Earlier available interpolated results for Cl XIV and K XVI are found to be underestimated by up to a factor of four, particularly for the forbidden transitions.  We believe the present set of complete results for radiative and excitation rates for transitions in CL XIV, K XVI and Ge XXIX will be useful for the modelling of a variety of plasmas, including astrophysical and  fusion. 

\section*{Acknowledgment}
KMA is thankful to  AWE Aldermaston for financial support.


\clearpage

\begin{figure*}
\includegraphics[scale=0.80,angle=90]{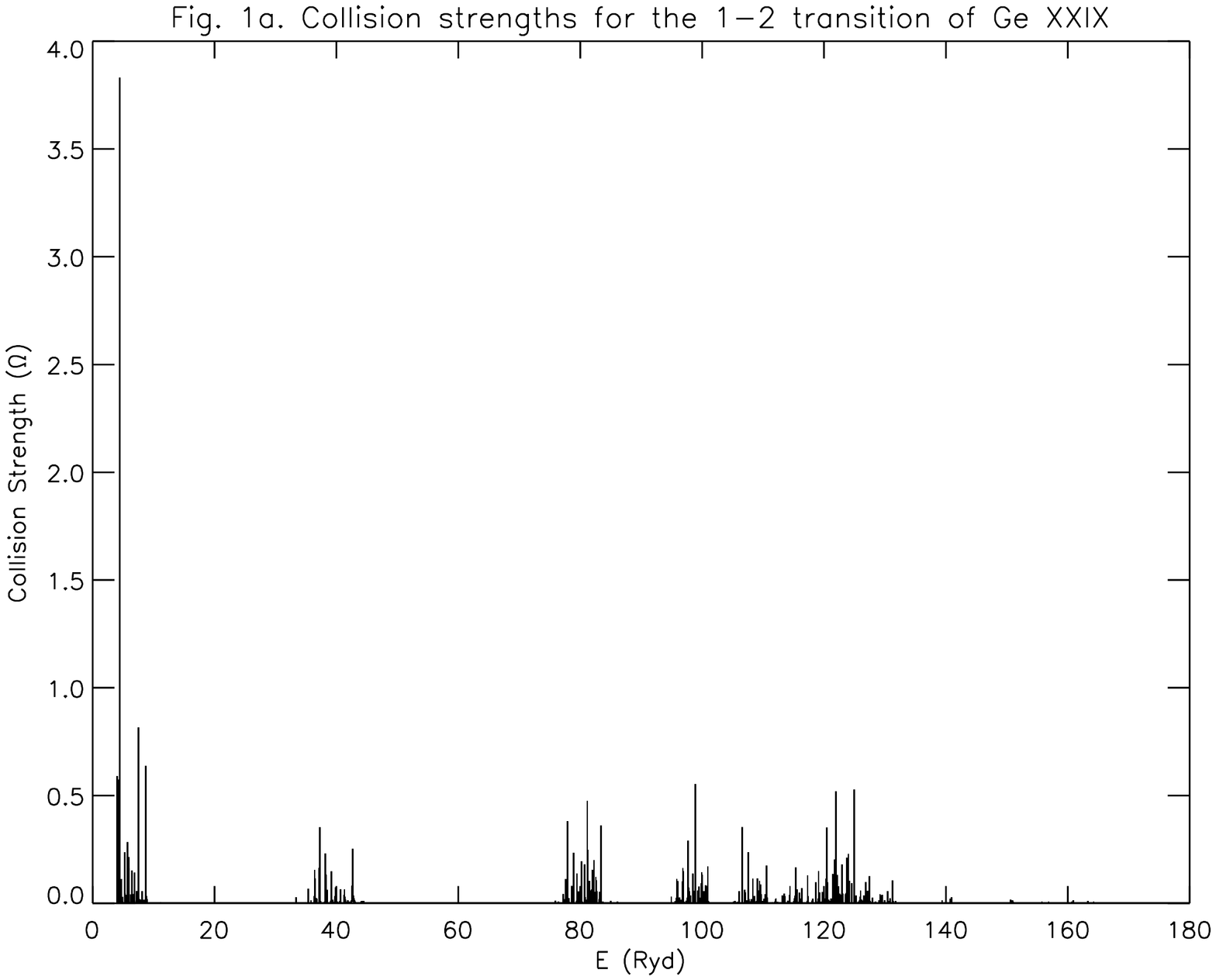}
\caption{{\bf  (a and b)} Collision strengths for the 2s$^2$ $^1$S$_0$ - 2s2p $^3$P$^o_0$   (1--2) transition of Ge XXIX.}
\end{figure*}

\setcounter{figure}{0}  
\begin{figure*}
\includegraphics[scale=0.80,angle=90]{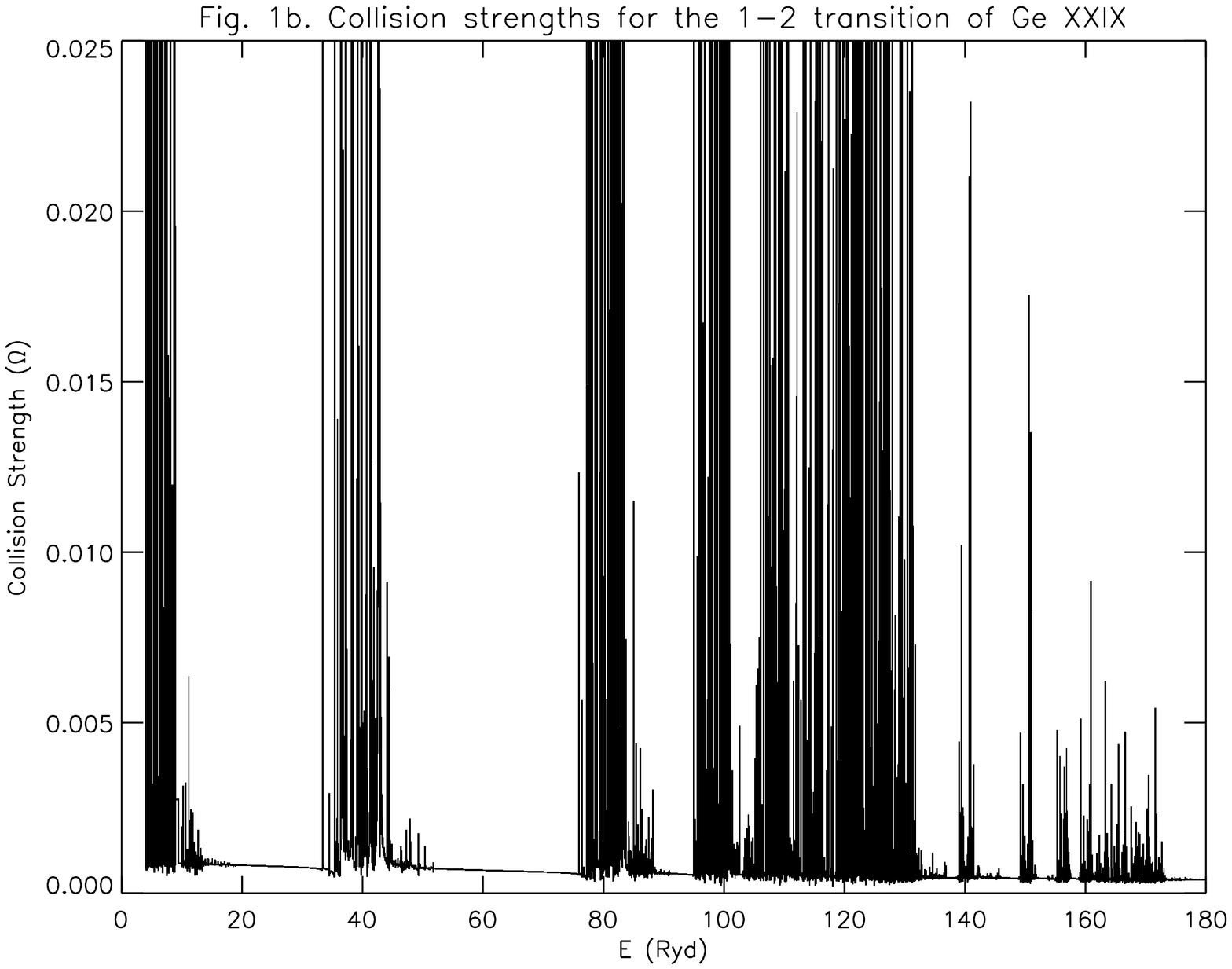}
\caption{{\bf  (a and b)} Collision strengths for the 2s$^2$ $^1$S$_0$ - 2s2p $^3$P$^o_0$   (1--2) transition of Ge XXIX.}
\end{figure*}

\setcounter{figure}{1}  
\begin{figure*}
\includegraphics[scale=0.80,angle=90]{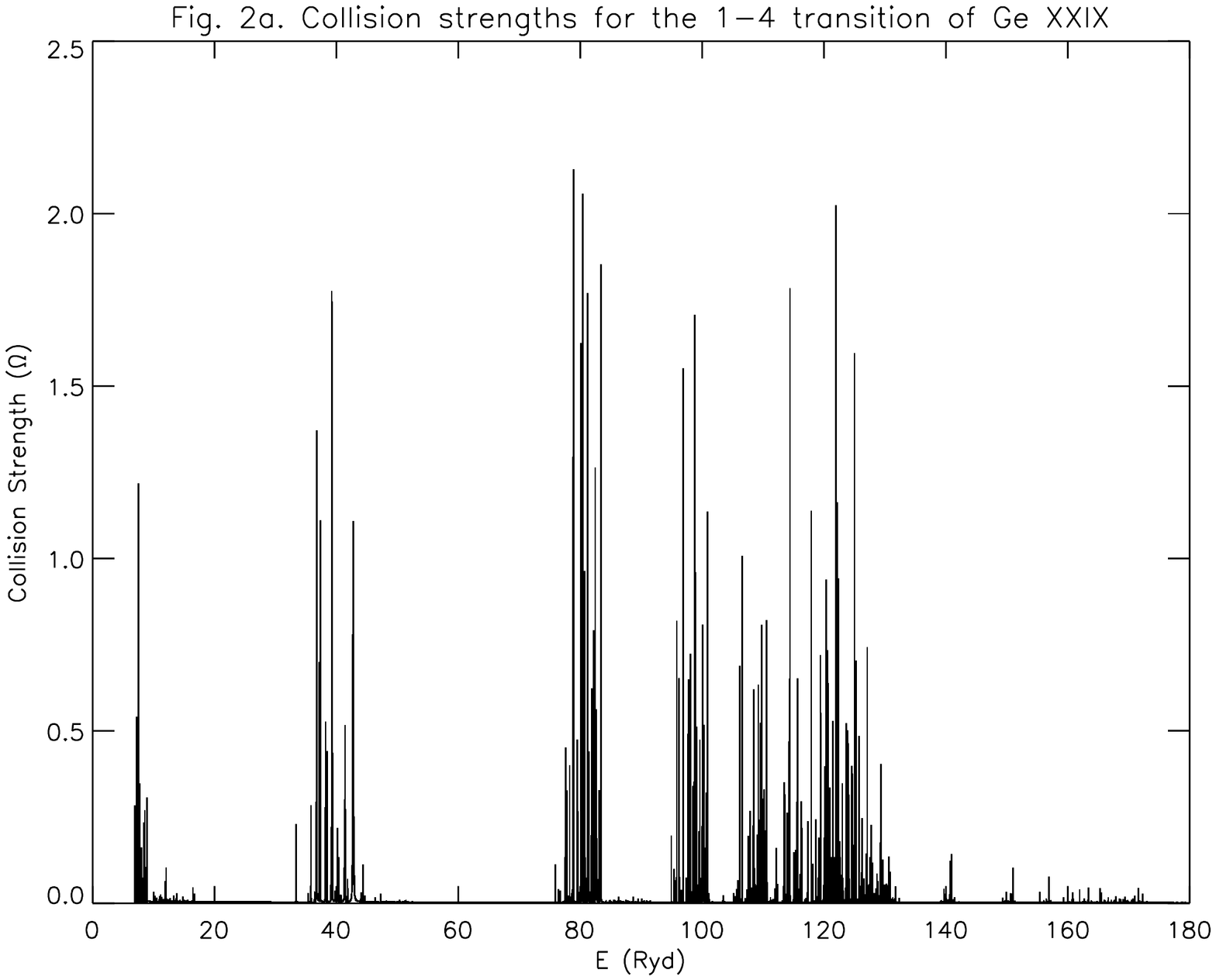}
\caption{{\bf  (a and b)} Collision strengths for the 2s$^2$ $^1$S$_0$ - 2s2p $^3$P$^o_2$ (1--4) transition of Ge XXIX.}
\end{figure*}

\setcounter{figure}{1}  
\begin{figure*}
\includegraphics[scale=0.80,angle=90]{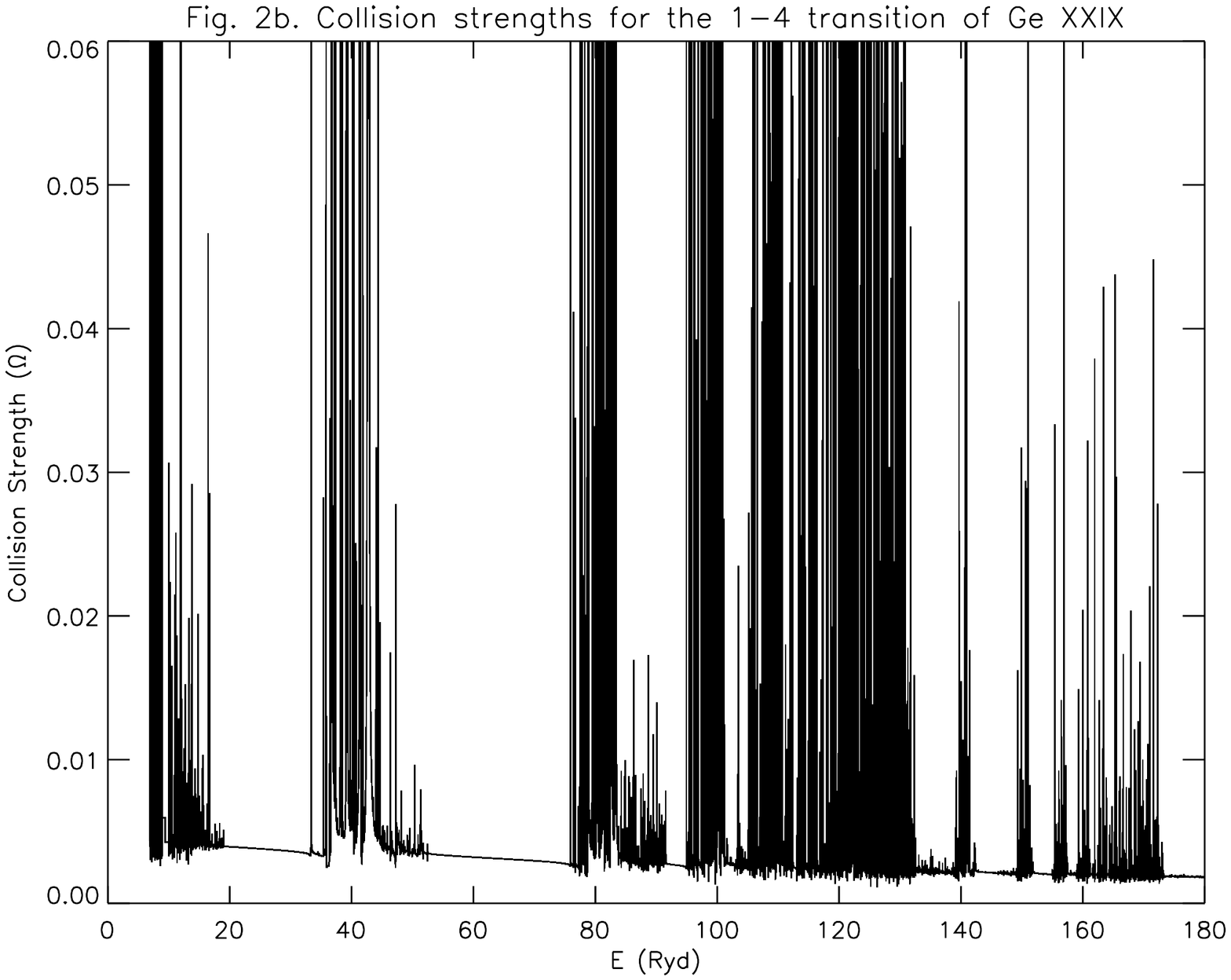}
\caption{{\bf  (a and b)} Collision strengths for the 2s$^2$ $^1$S$_0$ - 2s2p $^3$P$^o_2$ (1--4) transition of Ge XXIX.}
\end{figure*}

\setcounter{figure}{2}  
\begin{figure*}
\includegraphics[scale=0.80,angle=90]{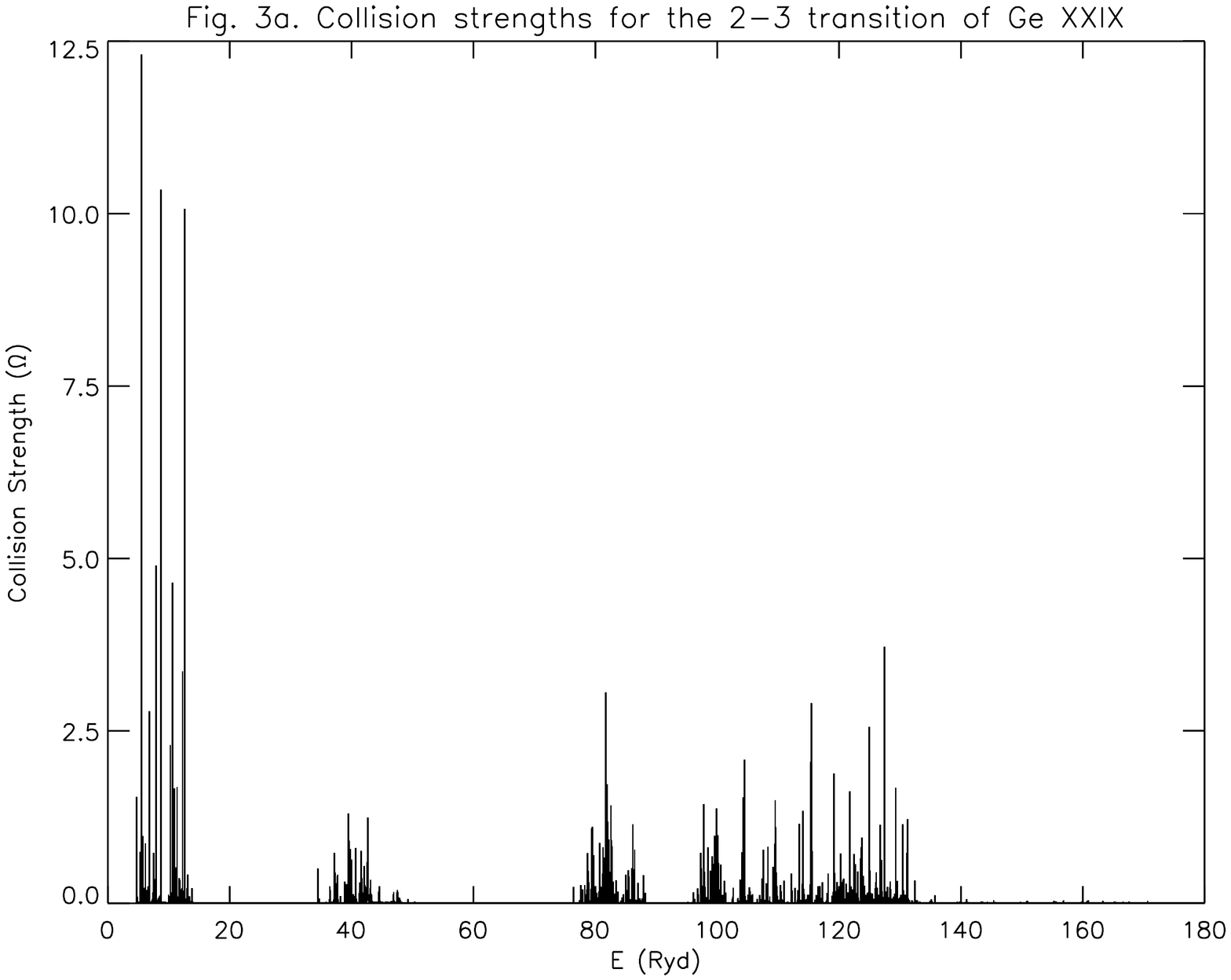}
\caption{{\bf  (a and b)} Collision strengths for the  2s2p $^3$P$^o_0$  - 2s2p $^3$P$^o_1$ (2--3) transition of Ge XXIX.}
\end{figure*}

\setcounter{figure}{2}  
\begin{figure*}
\includegraphics[scale=0.80,angle=90]{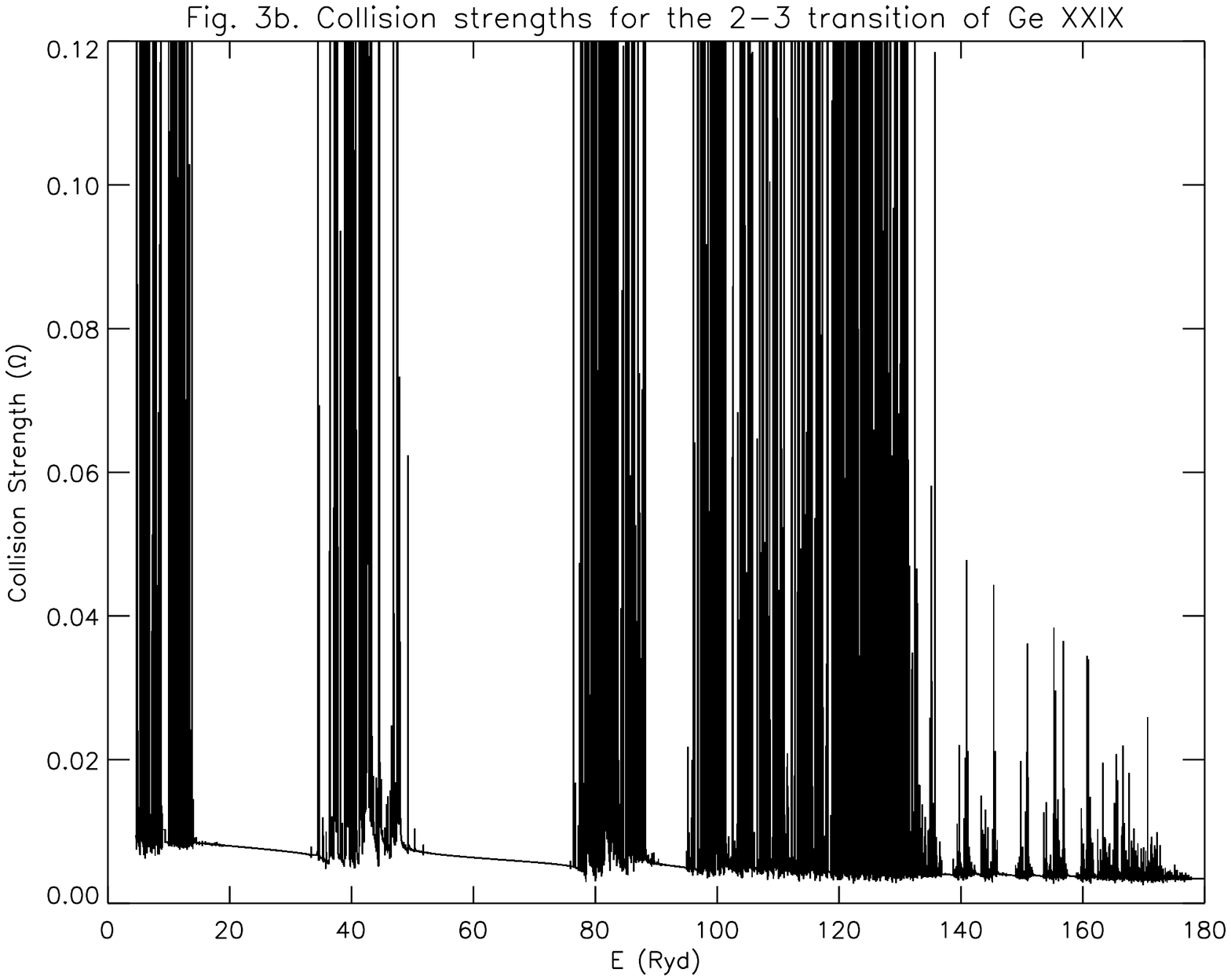}
\caption{{\bf  (a and b)} Collision strengths for the  2s2p $^3$P$^o_0$  - 2s2p $^3$P$^o_1$ (2--3) transition of Ge XXIX.}
\end{figure*}


\clearpage

\begin{table*} 
\caption{Energy levels (in Ryd) of Cl XIV and their lifetimes ($\tau$, s). $a{\pm}b \equiv a{\times}$10$^{{\pm}b}$.} 
\begin{tabular}{rllrrrrrrrl} \hline \hline
Index  & Configuration       & Level&  NIST   &   MBPT & MCHF        &   GRASP1	    & GRASP2	   &  FAC1      & FAC2	      &$\tau$ (s) \\
 \hline 
   1 &  2s$^2$    &  $^1$S$  _0$  &  0.0000  &   0.0000  &  0.0000  &	 0.0000   &   0.0000   &   0.0000  &	0.0000  &   ........	\\
   2 &  2s2p	  &  $^3$P$^o_0$  &  1.9493  &   1.9503  &  1.9481  &	 1.9503   &   1.9546   &   1.9628  &	1.9622  &   ........	\\
   3 &  2s2p	  &  $^3$P$^o_1$  &  2.0004  &   2.0014  &  1.9992  &	 2.0064   &   2.0055   &   2.0134  &	2.0128  &   6.052-07	\\
   4 &  2s2p	  &  $^3$P$^o_2$  &  2.1179  &   2.1190  &  2.1167  &	 2.1301   &   2.1222   &   2.1293  &	2.1288  &   3.464-02	\\
   5 &  2s2p	  &  $^1$P$^o_1$  &  3.8319  &   3.8284  &  3.8356  &	 3.9138   &   3.9108   &   3.9094  &	3.9046  &   1.102-10	\\
   6 &  2p$^2$    &  $^3$P$  _0$  &  5.1378  &   5.1388  &  5.1402  &	 5.1706   &   5.1729   &   5.1908  &	5.1900  &   1.500-10	\\
   7 &  2p$^2$    &  $^3$P$  _1$  &  5.2068  &   5.2079  &  5.2078  &	 5.2422   &   5.2408   &   5.2583  &	5.2575  &   1.449-10	\\
   8 &  2p$^2$    &  $^3$P$  _2$  &  5.3017  &   5.3032  &  5.3035  &	 5.3476   &   5.3361   &   5.3528  &	5.3521  &   1.415-10	\\
   9 &  2p$^2$    &  $^1$D$  _2$  &  5.8199  &   5.8158  &  5.8244  &	 5.9064   &   5.8973   &   5.9130  &	5.9090  &   5.677-10	\\
  10 &  2p$^2$    &  $^1$S$  _0$  &  7.1114  &   7.1076  &  7.1188  &	 7.2461   &   7.2471   &   7.2556  &	7.2529  &   7.051-11	\\
  11 &  2s3s	  &  $^3$S$  _1$  & 31.5180  &  31.5216  & 31.5240  &	31.5186   &  31.5033   &  31.5022  &   31.5019  &   1.584-12	\\
  12 &  2s3s	  &  $^1$S$  _0$  & 	     &  31.8883  & 31.8903  &	31.8868   &  31.8723   &  31.8854  &   31.8851  &   4.513-12	\\
  13 &  2s3p	  &  $^1$P$^o_1$  & 32.4147  &  32.4084  & 32.4107  &	32.4174   &  32.4020   &  32.4130  &   32.4122  &   7.494-13	\\
  14 &  2s3p	  &  $^3$P$^o_0$  & 	     &  	 & 32.4345  &	32.4337   &  32.4204   &  32.4321  &   32.4323  &   1.122-10	\\
  15 &  2s3p	  &  $^3$P$^o_1$  & 32.4766  &  32.4594  & 32.4605  &	32.4643   &  32.4480   &  32.4591  &   32.4589  &   2.019-12	\\
  16 &  2s3p	  &  $^3$P$^o_2$  & 	     &  	 & 32.4815  &	32.4841   &  32.4674   &  32.4784  &   32.4785  &   9.567-11	\\
  17 &  2s3d	  &  $^3$D$  _1$  & 32.9624  &  	 & 32.9635  &	32.9711   &  32.9528   &  32.9674  &   32.9652  &   2.960-13	\\
  18 &  2s3d	  &  $^3$D$  _2$  & 32.9715  &  32.9685  & 32.9695  &	32.9780   &  32.9587   &  32.9732  &   32.9710  &   2.980-13	\\
  19 &  2s3d	  &  $^3$D$  _3$  & 32.9778  &  	 & 32.9793  &	32.9885   &  32.9686   &  32.9829  &   32.9806  &   3.007-13	\\
  20 &  2s3d	  &  $^1$D$  _2$  & 33.3223  &  33.3151  & 33.3213  &	33.3611   &  33.3424   &  33.3559  &   33.3520  &   4.362-13	\\
  21 &  2p3s	  &  $^3$P$^o_0$  & 	     &  	 & 33.9707  &	33.9765   &  33.9640   &  33.9901  &   33.9903  &   2.066-12	\\
  22 &  2p3s	  &  $^3$P$^o_1$  & 	     &  34.0125  & 34.0143  &	34.0236   &  34.0086   &  34.0346  &   34.0347  &   2.005-12	\\
  23 &  2p3s	  &  $^3$P$^o_2$  & 	     &  	 & 34.1458  &	34.1613   &  34.1382   &  34.1629  &   34.1630  &   1.932-12	\\
  24 &  2p3s	  &  $^1$P$^o_1$  & 	     &  	 & 34.4507  &	34.4974   &  34.4773   &  34.5086  &   34.5047  &   1.548-12	\\
  25 &  2p3p	  &  $^1$P$  _1$  & 	     &  34.5548  & 	    &	34.5670   &  34.5529   &  34.5800  &   34.5800  &   1.245-12	\\
  26 &  2p3p	  &  $^3$D$  _1$  & 	     &  	 & 	    &	34.7028   &  34.6840   &  34.7098  &   34.7097  &   1.434-12	\\
  27 &  2p3p	  &  $^3$D$  _2$  & 34.8010  &  34.6993  & 	    &	34.7198   &  34.7020   &  34.7274  &   34.7272  &   1.906-12	\\
  28 &  2p3p	  &  $^3$D$  _3$  & 	     &  	 & 	    &	34.8474   &  34.8221   &  34.8462  &   34.8459  &   1.879-12	\\
  29 &  2p3p	  &  $^3$S$  _1$  & 	     &  34.9299  & 	    &	34.9529   &  34.9330   &  34.9609  &   34.9604  &   1.200-12	\\
  30 &  2p3p	  &  $^3$P$  _0$  & 	     &  	 & 	    &	34.9864   &  34.9730   &  35.0214  &   35.0201  &   1.124-12	\\
  31 &  2p3p	  &  $^3$P$  _1$  & 	     &  35.0462  & 	    &	35.0674   &  35.0477   &  35.0911  &   35.0901  &   1.136-12	\\
  32 &  2p3d	  &  $^3$F$^o_2$  & 	     &  	 & 35.0998  &	35.1111   &  35.0952   &  35.1303  &   35.1283  &   2.943-12	\\
  33 &  2p3p	  &  $^3$P$  _2$  & 35.0750  &  	 & 	    &	35.1128   &  35.0904   &  35.1364  &   35.1353  &   1.113-12	\\
  34 &  2p3d	  &  $^3$F$^o_3$  & 	     &  35.1761  & 35.1814  &	35.1992   &  35.1790   &  35.2181  &   35.2156  &   6.466-12	\\
  35 &  2p3d	  &  $^1$D$^o_2$  & 	     &  35.2343  & 	    &	35.2501   &  35.2279   &  35.2668  &   35.2661  &   7.737-13	\\
  36 &  2p3d	  &  $^3$F$^o_4$  & 	     &  	 & 35.2810  &	35.3058   &  35.2785   &  35.3132  &   35.3106  &   7.852-10	\\
  37 &  2p3p	  &  $^1$D$  _2$  & 	     &  35.3229  & 	    &	35.3762   &  35.3543   &  35.4072  &   35.4031  &   7.706-13	\\
  38 &  2p3d	  &  $^3$D$^o_1$  & 	     &  	 & 35.4474  &	35.4587   &  35.4393   &  35.4778  &   35.4780  &   2.435-13	\\
  39 &  2p3d	  &  $^3$D$^o_2$  & 35.4940  &  35.4740  & 35.4787  &	35.4923   &  35.4704   &  35.5121  &   35.5122  &   2.683-13	\\
  40 &  2p3d	  &  $^3$D$^o_3$  & 	     &  	 & 35.5393  &	35.5567   &  35.5303   &  35.5702  &   35.5704  &   2.403-13	\\
 \hline
\end{tabular} 
\end{table*} 

\clearpage  
\setcounter{table}{0} 

\begin{table*} 
\begin{tabular}{rllrrrrrrrl} \hline
 Index  & Configuration       & Level & NIST & MBPT & MCHF  & GRASP1 & GRASP2     &  FAC1      & FAC2        & $\tau$ (s) \\ 
\hline
  41 &  2p3d	  &  $^3$P$^o_2$  & 	     &  	 & 35.6233  &	35.6450   &  35.6192   &  35.6560  &   35.6558  &   3.799-13	\\
  42 &  2p3d	  &  $^3$P$^o_1$  & 	     &  35.6408  & 35.6450  &	35.6672   &  35.6420   &  35.6794  &   35.6792  &   4.088-13	\\
  43 &  2p3d	  &  $^3$P$^o_0$  & 	     &  	 & 35.6596  &	35.6804   &  35.6572   &  35.6949  &   35.6946  &   4.375-13	\\
  44 &  2p3p	  &  $^1$S$  _0$  & 	     &  35.7518  & 	    &	35.8426   &  35.8248   &  35.8827  &   35.8703  &   1.249-12	\\
  45 &  2p3d	  &  $^1$F$^o_3$  & 35.9400  &  35.9322  & 35.9497  &	36.0147   &  35.9882   &  36.0298  &   36.0231  &   1.872-13	\\
  46 &  2p3d	  &  $^1$P$^o_1$  & 	     &  36.0002  & 36.0081  &	36.0624   &  36.0386   &  36.0766  &   36.0743  &   3.137-13	\\
  47 &  2s4s	  &  $^3$S$  _1$  & 	     &  	 & 	    &	42.1793   &  42.1612   &  42.1662  &   42.1651  &   3.058-12	\\
  48 &  2s4s	  &  $^1$S$  _0$  & 	     &  	 & 	    &	42.3137   &  42.2962   &  42.3045  &   42.3017  &   3.618-12	\\
  49 &  2s4p	  &  $^3$P$^o_0$  &   &   & & 42.5337	  &  42.5165	&  42.5269  &  42.5269   &  8.031-12	\\
  50 &  2s4p	  &  $^3$P$^o_1$  &   &   & & 42.5387	  &  42.5210	&  42.5313  &  42.5313   &  6.176-12	\\
  51 &  2s4p	  &  $^3$P$^o_2$  &   &   & & 42.5548	  &  42.5361	&  42.5461  &  42.5462   &  8.187-12	\\
  52 &  2s4p	  &  $^1$P$^o_1$  &   &   & & 42.5857	  &  42.5669	&  42.5792  &  42.5768   &  1.024-12	\\
  53 &  2s4d	  &  $^3$D$  _1$  &   &   & & 42.7546	  &  42.7356	&  42.7438  &  42.7427   &  7.550-13	\\
  54 &  2s4d	  &  $^3$D$  _2$  &   &   & & 42.7571	  &  42.7377	&  42.7459  &  42.7448   &  7.577-13	\\
  55 &  2s4d	  &  $^3$D$  _3$  &   &   & & 42.7609	  &  42.7414	&  42.7495  &  42.7484   &  7.617-13	\\
  56 &  2s4d	  &  $^1$D$  _2$  &   &   & & 42.8827	  &  42.8635	&  42.8697  &  42.8677   &  8.569-13	\\
  57 &  2s4f	  &  $^3$F$^o_2$  &   &   & & 42.8910	  &  42.8719	&  42.8786  &  42.8764   &  1.795-12	\\
  58 &  2s4f	  &  $^3$F$^o_3$  &   &   & & 42.8922	  &  42.8728	&  42.8795  &  42.8773   &  1.795-12	\\
  59 &  2s4f	  &  $^3$F$^o_4$  &   &   & & 42.8939	  &  42.8743	&  42.8811  &  42.8788   &  1.796-12	\\
  60 &  2s4f	  &  $^1$F$^o_3$  &   &   & & 42.9244	  &  42.9050	&  42.9129  &  42.9104   &  1.826-12	\\
  61 &  2p4s	  &  $^3$P$^o_0$  &   &   & & 44.4581	  &  44.4426	&  44.4712  &  44.4713   &  3.501-12	\\
  62 &  2p4s	  &  $^3$P$^o_1$  &   &   & & 44.4817	  &  44.4655	&  44.4945  &  44.4933   &  3.186-12	\\
  63 &  2p4s	  &  $^3$P$^o_2$  &   &   & & 44.6452	  &  44.6193	&  44.6463  &  44.6464   &  3.156-12	\\
  64 &  2p4p	  &  $^3$D$  _1$  &   &   & & 44.7174	  &  44.7015	&  44.7329  &  44.7330   &  1.815-12	\\
  65 &  2p4s	  &  $^1$P$^o_1$  &   &   & & 44.7209	  &  44.6954	&  44.7249  &  44.7179   &  2.399-12	\\
  66 &  2p4p	  &  $^3$D$  _2$  &   &   & & 44.7955	  &  44.7773	&  44.8097  &  44.8094   &  2.032-12	\\
  67 &  2p4p	  &  $^3$P$  _1$  &   &   & & 44.7960	  &  44.7784	&  44.8112  &  44.8100   &  1.736-12	\\
  68 &  2p4p	  &  $^3$P$  _0$  &   &   & & 44.8683	  &  44.8518	&  44.8916  &  44.8887   &  1.913-12	\\
  69 &  2p4d	  &  $^3$F$^o_2$  &   &   & & 44.9356	  &  44.9187	&  44.9457  &  44.9439   &  2.478-12	\\
  70 &  2p4p	  &  $^3$D$  _3$  &   &   & & 44.9369	  &  44.9098	&  44.9499  &  44.9486   &  2.103-12	\\
  71 &  2p4p	  &  $^1$P$  _1$  &   &   & & 44.9387	  &  44.9135	&  44.9395  &  44.9396   &  1.680-12	\\
  72 &  2p4d	  &  $^3$F$^o_3$  &   &   & & 44.9982	  &  44.9794	&  45.0127  &  45.0114   &  1.708-12	\\
  73 &  2p4p	  &  $^3$P$  _2$  &   &   & & 45.0016	  &  44.9760	&  45.0104  &  45.0070   &  1.874-12	\\
  74 &  2p4p	  &  $^3$S$  _1$  &   &   & & 45.0019	  &  44.9762	&  45.0105  &  45.0088   &  1.756-12	\\
  75 &  2p4d	  &  $^3$D$^o_2$  &   &   & & 45.0139	  &  44.9958	&  45.0265  &  45.0256   &  9.595-13	\\
  76 &  2p4d	  &  $^3$D$^o_1$  &   &   & & 45.0634	  &  45.0454	&  45.0748  &  45.0741   &  6.067-13	\\
  77 &  2p4f	  &  $^3$G$  _3$  &   &   & & 45.0846	  &  45.0671	&  45.0942  &  45.0910   &  1.816-12	\\
  78 &  2p4p	  &  $^1$D$  _2$  &   &   & & 45.0857	  &  45.0636	&  45.1223  &  45.1193   &  1.672-12	\\
  79 &  2p4f	  &  $^3$F$  _3$  &   &   & & 45.0937	  &  45.0761	&  45.1034  &  45.1034   &  1.817-12	\\
  80 &  2p4f	  &  $^3$G$  _4$  &   &   & & 45.0975	  &  45.0799	&  45.1074  &  45.1040   &  1.867-12	\\
 \hline
\end{tabular} 
\end{table*} 

\clearpage  
\setcounter{table}{0} 

\begin{table*} 
\begin{tabular}{rllrrrrrrrl} \hline
 Index  & Configuration       & Level & NIST & MBPT & MCHF  & GRASP1 & GRASP2     &  FAC1      & FAC2        & $\tau$ (s) \\ 
\hline
  81 &  2p4f	  &  $^3$F$  _2$  &   &   & & 45.1092	  &  45.0878	&  45.0974  &  45.0961   &  1.671-12	\\
  82 &  2p4d	  &  $^3$F$^o_4$  &   &   & & 45.1288	  &  45.1013	&  45.1312  &  45.1295   &  3.697-12	\\
  83 &  2p4d	  &  $^1$D$^o_2$  &   &   & & 45.1504	  &  45.1242	&  45.1536  &  45.1528   &  9.341-13	\\
  84 &  2p4d	  &  $^3$D$^o_3$  &   &   & & 45.1926	  &  45.1656	&  45.1942  &  45.1940   &  6.453-13	\\
  85 &  2p4d	  &  $^3$P$^o_2$  &   &   & & 45.2271	  &  45.2000	&  45.2286  &  45.2272   &  7.548-13	\\
  86 &  2p4d	  &  $^3$P$^o_1$  &   &   & & 45.2371	  &  45.2103	&  45.2391  &  45.2370   &  8.236-13	\\
  87 &  2p4d	  &  $^3$P$^o_0$  &   &   & & 45.2437	  &  45.2178	&  45.2467  &  45.2441   &  9.210-13	\\
  88 &  2p4f	  &  $^1$F$  _3$  &   &   & & 45.2561	  &  45.2288	&  45.2545  &  45.2524   &  1.814-12	\\
  89 &  2p4f	  &  $^3$F$  _4$  &   &   & & 45.2639	  &  45.2366	&  45.2623  &  45.2599   &  1.838-12	\\
  90 &  2p4p	  &  $^1$S$  _0$  &   &   & & 45.2835	  &  45.2594	&  45.2920  &  45.2774   &  2.414-12	\\
  91 &  2p4f	  &  $^3$G$  _5$  &   &   & & 45.2926	  &  45.2648	&  45.2989  &  45.2842   &  1.830-12	\\
  92 &  2p4f	  &  $^3$D$  _3$  &   &   & & 45.2929	  &  45.2661	&  45.3094  &  45.2921   &  1.807-12	\\
  93 &  2p4f	  &  $^3$D$  _2$  &   &   & & 45.2993	  &  45.2725	&  45.2903  &  45.2988   &  1.807-12	\\
  94 &  2p4f	  &  $^1$G$  _4$  &   &   & & 45.3119	  &  45.2842	&  45.3107  &  45.3042   &  1.955-12	\\
  95 &  2p4f	  &  $^3$D$  _1$  &   &   & & 45.3289	  &  45.3020	&  45.3280  &  45.3281   &  1.801-12	\\
  96 &  2p4f	  &  $^1$D$  _2$  &   &   & & 45.3444	  &  45.3174	&  45.3441  &  45.3438   &  1.807-12	\\
  97 &  2p4d	  &  $^1$F$^o_3$  &   &   & & 45.3605	  &  45.3338	&  45.3604  &  45.3527   &  4.311-13	\\
  98 &  2p4d	  &  $^1$P$^o_1$  &   &   & & 45.3767	  &  45.3510	&  45.3775  &  45.3721   &  6.594-13	\\
 \hline	
\end{tabular}

\begin{flushleft}
{\small
NIST: {\tt http://www.nist.gov/pml/data/asd.cfm} \\
MBPT: Gu \cite{gu} for the lowest 10 levels and Safronova {\em et al} \cite{uis4} for the remaining levels \\
MCHF: Calculations of C Froese Fischer (2009) available at {\tt http://nlte.nist.gov/MCHF/view.html} \\
GRASP1: Coulomb energies \\
GRASP2: QED corrected energies \\
FAC1: Energies from the FAC for 98 level calculations\\
FAC2: Energies from the FAC for 166 level calculations \\
}
\end{flushleft}
\end{table*} 

\clearpage  
\setcounter{table}{1} 
\begin{table*} 
\caption{Energy levels (in Ryd) of K XVI and their lifetimes ($\tau$, s). $a{\pm}b \equiv a{\times}$10$^{{\pm}b}$.} 
\begin{tabular}{rllrrrrrrl} \hline \hline
Index  & Configuration       & Level&  NIST   &   MBPT       &   GRASP1	    & GRASP2	   &  FAC1      & FAC2	      &$\tau$ (s) \\
 \hline 
   1 &  2s$^2$    &  $^1$S$  _0$  &   0.0000  &   0.0000     &   0.0000    &   0.0000	&   0.0000  &   0.0000   &   ........	 \\
   2 &  2s2p	  &  $^3$P$^o_0$  &   2.2191  &   2.2198     &   2.2183    &   2.2241   &   2.2325  &   2.2319   &   ........	 \\
   3 &  2s2p	  &  $^3$P$^o_1$  &   2.3011  &   2.3023     &   2.3079    &   2.3064   &   2.3143  &   2.3137   &   2.421-07	 \\
   4 &  2s2p	  &  $^3$P$^o_2$  &   2.4977  &   2.4994     &   2.5141    &   2.5023   &   2.5092  &   2.5087   &   7.442-03	 \\
   5 &  2s2p	  &  $^1$P$^o_1$  &   4.4178  &   4.4157     &   4.5028    &   4.4974   &   4.4959  &   4.4911   &   9.267-11	 \\
   6 &  2p$^2$    &  $^3$P$  _0$  &   5.8933  &   5.8947     &   5.9272    &   5.9301   &   5.9481  &   5.9473   &   1.285-10	 \\
   7 &  2p$^2$    &  $^3$P$  _1$  &   6.0162  &   6.0174     &   6.0539    &   6.0508   &   6.0681  &   6.0674   &   1.217-10	 \\
   8 &  2p$^2$    &  $^3$P$  _2$  &   6.1647  &   6.1663     &   6.2176    &   6.2005   &   6.2171  &   6.2162   &   1.195-10	 \\
   9 &  2p$^2$    &  $^1$D$  _2$  &   6.7675  &   6.7641     &   6.8594    &   6.8442   &   6.8596  &   6.8557   &   4.047-10	 \\
  10 &  2p$^2$    &  $^1$S$  _0$  &   8.2266  &   8.2241     &   8.3640    &   8.3631   &   8.3716  &   8.3689   &   5.888-11	 \\
  11 &  2s3s	  &  $^3$S$  _1$  &  	      &  40.5331     &  40.5355    &  40.5139   &  40.5127  &  40.5125   &   9.796-13	 \\
  12 &  2s3s	  &  $^1$S$  _0$  &  	      &  40.9544     &  40.9582    &  40.9378   &  40.9509  &  40.9507   &   2.792-12	 \\
  13 &  2s3p	  &  $^1$P$^o_1$  &  41.5890  &  41.5561     &  41.5694    &  41.5481   &  41.5593  &  41.5587   &   5.487-13	 \\
  14 &  2s3p	  &  $^3$P$^o_0$  &  	      & 	     &  41.5822    &  41.5635   &  41.5751  &  41.5753   &   6.808-11	 \\
  15 &  2s3p	  &  $^3$P$^o_1$  &  	      &  41.6281     &  41.6410    &  41.6174   &  41.6282  &  41.6279   &   7.904-13	 \\
  16 &  2s3p	  &  $^3$P$^o_2$  &  	      & 	     &  41.6656    &  41.6420   &  41.6526  &  41.6528   &   5.444-11	 \\
  17 &  2s3d	  &  $^3$D$  _1$  &           & 	     &  42.2243    &  42.1983   &  42.2126  &  42.2104   &   1.766-13	 \\
  18 &  2s3d	  &  $^3$D$  _2$  &  42.2460  &  42.2187     &  42.2359    &  42.2084   &  42.2225  &  42.2203   &   1.783-13	 \\
  19 &  2s3d	  &  $^3$D$  _3$  &  42.2550  & 	     &  42.2533    &  42.2249   &  42.2389  &  42.2366   &   1.805-13	 \\
  20 &  2s3d	  &  $^1$D$  _2$  &  42.6380  &  42.6222     &  42.6760    &  42.6494   &  42.6626  &  42.6586   &   2.584-13	 \\
  21 &  2p3s	  &  $^3$P$^o_0$  &  	      & 	     &  43.3296    &  43.3118   &  43.3380  &  43.3381   &   1.301-12	 \\
  22 &  2p3s	  &  $^3$P$^o_1$  &  	      &  43.3811     &  43.3987    &  43.3780   &  43.4042  &  43.4041   &   1.239-12	 \\
  23 &  2p3s	  &  $^3$P$^o_2$  &  	      & 	     &  43.6320    &  43.5989   &  43.6234  &  43.6235   &   1.182-12	 \\
  24 &  2p3s	  &  $^1$P$^o_1$  &  	      &  43.9353     &  43.9969    &  43.9674   &  43.9982  &  43.9945   &   9.444-13	 \\
  25 &  2p3p	  &  $^3$D$  _1$  &  	      & 	     &  44.0149    &  43.9961   &  44.0234  &  44.0234   &   8.080-13	 \\
  26 &  2p3p	  &  $^1$P$  _1$  &  	      &  43.9971     &  44.2220    &  44.1957   &  44.2220  &  44.2220   &   7.785-13	 \\
  27 &  2p3p	  &  $^3$D$  _2$  &  	      &  44.2017     &  44.2280    &  44.2029   &  44.2288  &  44.2285   &   1.085-12	 \\
  28 &  2p3p	  &  $^3$D$  _3$  &  44.4200  & 	     &  44.4434    &  44.4068   &  44.4306  &  44.4304   &   1.074-12	 \\
  29 &  2p3p	  &  $^3$P$  _0$  &  	      & 	     &  44.5099    &  44.4913   &  44.5400  &  44.5386   &   6.573-13	 \\
  30 &  2p3p	  &  $^3$S$  _1$  &  	      &  44.4942     &  44.5255    &  44.4961   &  44.5254  &  44.5250   &   7.008-13	 \\
  31 &  2p3d	  &  $^3$F$^o_2$  &  	      & 	     &  44.6584    &  44.6356   &  44.6707  &  44.6688   &   1.714-12	 \\
  32 &  2p3p	  &  $^3$P$  _1$  &  	      &  44.6337     &  44.6651    &  44.6355   &  44.6764  &  44.6754   &   6.717-13	 \\
  33 &  2p3p	  &  $^3$P$  _2$  &  	      & 	     &  44.7225    &  44.6903   &  44.7360  &  44.7348   &   6.444-13	 \\
  34 &  2p3d	  &  $^3$F$^o_3$  &  	      &  44.7634     &  44.7947    &  44.7663   &  44.8055  &  44.8031   &   2.114-12	 \\
  35 &  2p3d	  &  $^1$D$^o_2$  &  	      &  44.8466     &  44.8706    &  44.8404   &  44.8797  &  44.8791   &   4.040-13	 \\
  36 &  2p3d	  &  $^3$F$^o_4$  &  	      & 	     &  44.9790    &  44.9396   &  44.9739  &  44.9713   &   4.452-10	 \\
  37 &  2p3p	  &  $^1$D$  _2$  &  	      &  44.9660     &  45.0283    &  44.9959   &  45.0481  &  45.0442   &   4.606-13	 \\
  38 &  2p3d	  &  $^3$D$^o_1$  &  	      & 	     &  45.0766    &  45.0495   &  45.0877  &  45.0878   &   1.478-13	 \\
  39 &  2p3d	  &  $^3$D$^o_2$  &  45.1810  &  45.1199     &  45.1492    &  45.1164   &  45.1570  &  45.1570   &   1.758-13	 \\
  40 &  2p3d	  &  $^3$D$^o_3$  &  45.2500  & 	     &  45.2543    &  45.2159   &  45.2554  &  45.2555   &   1.464-13	 \\
 \hline
\end{tabular} 
\end{table*} 

\clearpage  
\setcounter{table}{1} 
\begin{table*} 

\begin{tabular}{rllrrrrrrl} \hline
 Index  & Configuration       & Level & NIST & MBPT & GRASP1 & GRASP2     &  FAC1      & FAC2        & $\tau$ (s) \\ 
\hline
  41 &  2p3d	  &  $^3$P$^o_2$  &  	      & 	     &  45.3568    &  45.3189   &  45.3556  &  45.3555   &   2.148-13	 \\
  42 &  2p3d	  &  $^3$P$^o_1$  &  	      &  45.3443     &  45.3819    &  45.3452   &  45.3826  &  45.3825   &   2.354-13	 \\
  43 &  2p3d	  &  $^3$P$^o_0$  &  	      & 	     &  45.3986    &  45.3652   &  45.4026  &  45.4024   &   2.595-13	 \\
  44 &  2p3p	  &  $^1$S$  _0$  &  	      &  45.4588     &  45.5575    &  45.5310   &  45.5886  &  45.5764   &   7.079-13	 \\
  45 &  2p3d	  &  $^1$F$^o_3$  &  45.7120  &  45.6848     &  45.7785    &  45.7397   &  45.7808  &  45.7743   &   1.112-13	 \\
  46 &  2p3d	  &  $^1$P$^o_1$  &  	      &  45.7565     &  45.8293    &  45.7945   &  45.8321  &  45.8299   &   1.856-13	 \\
  47 &  2s4s	  &  $^3$S$  _1$  &  	      & 	     &  54.2989    &  54.2733   &  54.2781  &  54.2770   &   1.873-12	 \\
  48 &  2s4s	  &  $^1$S$  _0$  &  	      & 	     &  54.4509    &  54.4262   &  54.4343  &  54.4316   &   2.271-12	 \\
  49 &  2s4p	  &  $^3$P$^o_0$  &   &   &  54.7057	 &  54.6814    &  54.6916  &  54.6917	&   4.641-12   \\
  50 &  2s4p	  &  $^3$P$^o_1$  &   &   &  54.7129	 &  54.6879    &  54.6981  &  54.6980	&   3.098-12   \\
  51 &  2s4p	  &  $^3$P$^o_2$  &   &   &  54.7407	 &  54.7143    &  54.7239  &  54.7241	&   4.752-12   \\
  52 &  2s4p	  &  $^1$P$^o_1$  &   &   &  54.7740	 &  54.7474    &  54.7592  &  54.7569	&   6.286-13   \\
  53 &  2s4d	  &  $^3$D$  _1$  &   &   &  54.9710	 &  54.9440    &  54.9520  &  54.9509	&   4.507-13   \\
  54 &  2s4d	  &  $^3$D$  _2$  &   &   &  54.9751	 &  54.9476    &  54.9555  &  54.9544	&   4.528-13   \\
  55 &  2s4d	  &  $^3$D$  _3$  &   &   &  54.9816	 &  54.9538    &  54.9617  &  54.9606	&   4.561-13   \\
  56 &  2s4d	  &  $^1$D$  _2$  &   &   &  55.1184	 &  55.0911    &  55.0971  &  55.0950	&   5.110-13   \\
  57 &  2s4f	  &  $^3$F$^o_2$  &   &   &  55.1308	 &  55.1036    &  55.1103  &  55.1079	&   1.058-12   \\
  58 &  2s4f	  &  $^3$F$^o_3$  &   &   &  55.1328	 &  55.1052    &  55.1118  &  55.1094	&   1.058-12   \\
  59 &  2s4f	  &  $^3$F$^o_4$  &   &   &  55.1358	 &  55.1079    &  55.1145  &  55.1121	&   1.059-12   \\
  60 &  2s4f	  &  $^1$F$^o_3$  &   &   &  55.1696	 &  55.1420    &  55.1497  &  55.1471	&   1.078-12   \\
  61 &  2p4s	  &  $^3$P$^o_0$  &   &   &  56.8894	 &  56.8672    &  56.8958  &  56.8959	&   2.197-12   \\
  62 &  2p4s	  &  $^3$P$^o_1$  &   &   &  56.9187	 &  56.8958    &  56.9249  &  56.9236	&   1.951-12   \\
  63 &  2p4p	  &  $^3$D$  _1$  &   &   &  57.1880	 &  57.1655    &  57.1970  &  57.1971	&   1.099-12   \\
  64 &  2p4s	  &  $^3$P$^o_2$  &   &   &  57.1951	 &  57.1579    &  57.1845  &  57.1847	&   1.883-12   \\
  65 &  2p4s	  &  $^1$P$^o_1$  &   &   &  57.2776	 &  57.2408    &  57.2697  &  57.2637	&   1.512-12   \\
  66 &  2p4p	  &  $^3$P$  _1$  &   &   &  57.2945	 &  57.2703    &  57.3039  &  57.3025	&   1.048-12   \\
  67 &  2p4p	  &  $^3$D$  _2$  &   &   &  57.2962	 &  57.2709    &  57.3040  &  57.3035	&   1.177-12   \\
  68 &  2p4p	  &  $^3$P$  _0$  &   &   &  57.3702	 &  57.3474    &  57.3877  &  57.3841	&   1.126-12   \\
  69 &  2p4d	  &  $^3$F$^o_2$  &   &   &  57.4495	 &  57.4253    &  57.4565  &  57.4552	&   1.452-12   \\
  70 &  2p4p	  &  $^1$P$  _1$  &   &   &  57.5335	 &  57.4964    &  57.5276  &  57.5267	&   9.950-13   \\
  71 &  2p4d	  &  $^3$F$^o_3$  &   &   &  57.5359	 &  57.5096    &  57.5406  &  57.5389	&   8.382-13   \\
  72 &  2p4p	  &  $^3$D$  _3$  &   &   &  57.5417	 &  57.5028    &  57.5320  &  57.5322	&   1.234-12   \\
  73 &  2p4d	  &  $^3$D$^o_2$  &   &   &  57.5490	 &  57.5237    &  57.5543  &  57.5535	&   5.351-13   \\
  74 &  2p4p	  &  $^3$P$  _2$  &   &   &  57.6018	 &  57.5651    &  57.6009  &  57.5996	&   1.096-12   \\
  75 &  2p4d	  &  $^3$D$^o_1$  &   &   &  57.6025	 &  57.5771    &  57.6065  &  57.6057	&   3.680-13   \\
  76 &  2p4p	  &  $^3$S$  _1$  &   &   &  57.6076	 &  57.5706    &  57.6040  &  57.6006	&   1.040-12   \\
  77 &  2p4f	  &  $^3$G$  _3$  &   &   &  57.6251	 &  57.6001    &  57.6273  &  57.6241	&   1.069-12   \\
  78 &  2p4f	  &  $^3$F$  _3$  &   &   &  57.6390	 &  57.6139    &  57.6413  &  57.6414	&   1.069-12   \\
  79 &  2p4f	  &  $^3$F$  _2$  &   &   &  57.6402	 &  57.6143    &  57.6424  &  57.6425	&   1.080-12   \\
  80 &  2p4f	  &  $^3$G$  _4$  &   &   &  57.6422	 &  57.6171    &  57.6447  &  57.6414	&   1.098-12   \\
 \hline
\end{tabular} 
\end{table*} 

\clearpage  
\setcounter{table}{1} 
\begin{table*} 

\begin{tabular}{rllrrrrrrl} \hline
 Index  & Configuration       & Level & NIST & MBPT & GRASP1 & GRASP2     &  FAC1      & FAC2        & $\tau$ (s) \\ 
\hline
  81 &  2p4p	  &  $^1$D$  _2$  &   &   &  57.7208	 &  57.6837    &  57.7233  &  57.7192	&   9.315-13   \\
  82 &  2p4d	  &  $^3$F$^o_4$  &   &   &  57.7661	 &  57.7264    &  57.7558  &  57.7542	&   2.166-12   \\
  83 &  2p4d	  &  $^1$D$^o_2$  &   &   &  57.7837	 &  57.7452    &  57.7742  &  57.7735	&   6.175-13   \\
  84 &  2p4d	  &  $^3$D$^o_3$  &   &   &  57.8344	 &  57.7951    &  57.8234  &  57.8231	&   4.084-13   \\
  85 &  2p4d	  &  $^3$P$^o_2$  &   &   &  57.8761	 &  57.8368    &  57.8650  &  57.8639	&   4.466-13   \\
  86 &  2p4d	  &  $^3$P$^o_1$  &   &   &  57.8864	 &  57.8476    &  57.8760  &  57.8743	&   4.830-13   \\
  87 &  2p4d	  &  $^3$P$^o_0$  &   &   &  57.8941	 &  57.8567    &  57.8852  &  57.8830	&   5.481-13   \\
  88 &  2p4f	  &  $^1$F$  _3$  &   &   &  57.9122	 &  57.8727    &  57.8981  &  57.8964	&   1.067-12   \\
  89 &  2p4p	  &  $^1$S$  _0$  &   &   &  57.9207	 &  57.8857    &  57.9342  &  57.9065	&   1.358-12   \\
  90 &  2p4f	  &  $^3$F$  _4$  &   &   &  57.9223	 &  57.8827    &  57.9081  &  57.9062	&   1.079-12   \\
  91 &  2p4f	  &  $^3$D$  _3$  &   &   &  57.9533	 &  57.9142    &  57.9398  &  57.9399	&   1.064-12   \\
  92 &  2p4f	  &  $^3$G$  _5$  &   &   &  57.9567	 &  57.9165    &  57.9417  &  57.9362	&   1.076-12   \\
  93 &  2p4f	  &  $^3$D$  _2$  &   &   &  57.9588	 &  57.9198    &  57.9458  &  57.9458	&   1.065-12   \\
  94 &  2p4f	  &  $^1$G$  _4$  &   &   &  57.9778	 &  57.9377    &  57.9639  &  57.9581	&   1.138-12   \\
  95 &  2p4f	  &  $^3$D$  _1$  &   &   &  57.9965	 &  57.9577    &  57.9834  &  57.9835	&   1.060-12   \\
  96 &  2p4f	  &  $^1$D$  _2$  &   &   &  58.0150	 &  57.9760    &  58.0023  &  58.0020	&   1.065-12   \\
  97 &  2p4d	  &  $^1$F$^o_3$  &   &   &  58.0209	 &  57.9818    &  58.0081  &  58.0013	&   2.652-13   \\
  98 &  2p4d	  &  $^1$P$^o_1$  &   &   &  58.0413	 &  58.0037    &  58.0299  &  58.0251	&   3.988-13   \\
 \hline	
\end{tabular}

\begin{flushleft}
{\small
NIST: {\tt http://www.nist.gov/pml/data/asd.cfm} \\
MBPT: Gu \cite{gu} for the lowest 10 levels and Safronova {\em et al} \cite{uis4} for the remaining levels \\
GRASP1: Coulomb energies \\
GRASP2: QED corrected energies \\
FAC1: Energies from the FAC for 98 level calculations\\
FAC2: Energies from the FAC for 166 level calculations \\
}
\end{flushleft}

\end{table*} 

\clearpage  
\setcounter{table}{2} 
\begin{table*}                                                                                                
\caption{Energy levels (in Ryd) of Ge XXIX and their lifetimes ($\tau$, s). $a{\pm}b \equiv a{\times}$10$^{{\pm}b}$.} 
\begin{tabular}{rllrrrrrrl} \hline \hline
Index  & Configuration       & Level&  NIST   &   MBPT       &   GRASP1	    & GRASP2	   &  FAC1      & FAC2	      &$\tau$ (s) \\
 \hline 
   1 &  2s$^2$    &  $^1$S$  _0$  & 0.0000  &   0.0000     &	 0.0000   &    0.0000	&   0.0000   &    0.0000   &   ........    \\
   2 &  2s2p	  &  $^3$P$^o_0$  & 	    &   4.0218     &	 4.0076   &    4.0261	&   4.0356   &    4.0349   &   ........    \\
   3 &  2s2p	  &  $^3$P$^o_1$  & 4.5709  &   4.5745     &	 4.5941   &    4.5830	&   4.5913   &    4.5903   &   4.912-09    \\
   4 &  2s2p	  &  $^3$P$^o_2$  & 6.8006  &   6.8054     &	 6.8820   &    6.8037	&   6.8104   &    6.8099   &   5.524-06    \\
   5 &  2s2p	  &  $^1$P$^o_1$  & 9.8089  &   9.8159     &	 9.9459   &    9.8887	&   9.8874   &    9.8828   &   2.720-11    \\
   6 &  2p$^2$    &  $^3$P$  _0$  & 	    &  11.3155     &	11.3508   &   11.3709	&  11.3898   &   11.3883   &   5.842-11    \\
   7 &  2p$^2$    &  $^3$P$  _1$  & 	    &  13.2583     &	13.3369   &   13.2925	&  13.3102   &   13.3092   &   3.607-11    \\
   8 &  2p$^2$    &  $^3$P$  _2$  & 	    &  13.8769     &	14.0243   &   13.9304	&  13.9470   &   13.9446   &   4.777-11    \\
   9 &  2p$^2$    &  $^1$D$  _2$  & 	    &  16.5374     &	16.7319   &   16.5921	&  16.6061   &   16.6036   &   3.584-11    \\
  10 &  2p$^2$    &  $^1$S$  _0$  & 	    &  18.8739     &	19.0658   &   18.9956	&  19.0050   &   19.0025   &   1.650-11    \\
  11 &  2s3s	  &  $^3$S$  _1$  & 	    &              &   127.6944   &  127.5775	& 127.5759   &  127.5758   &   1.109-13    \\
  12 &  2s3s	  &  $^1$S$  _0$  & 	    &              &   128.4947   &  128.3840	& 128.3968   &  128.3965   &   2.659-13    \\
  13 &  2s3p	  &  $^3$P$^o_0$  & 	    &              &   129.6281   &  129.5241	& 129.5353   &  129.5354   &   1.260-11    \\
  14 &  2s3p	  &  $^3$P$^o_1$  & 	    &              &   129.6476   &  129.5327	& 129.5436   &  129.5432   &   8.841-14    \\
  15 &  2s3p	  &  $^1$P$^o_1$  & 	    &              &   130.4135   &  130.2782	& 130.2887   &  130.2880   &   4.847-14    \\
  16 &  2s3p	  &  $^3$P$^o_2$  & 	    &              &   130.4565   &  130.3238	& 130.3324   &  130.3326   &   5.071-12    \\
  17 &  2s3d	  &  $^3$D$  _1$  & 	    &              &   131.4705   &  131.3240	& 131.3359   &  131.3336   &   1.707-14    \\
  18 &  2s3d	  &  $^3$D$  _2$  & 	    &              &   131.5755   &  131.4205	& 131.4318   &  131.4294   &   1.786-14    \\
  19 &  2s3d	  &  $^3$D$  _3$  & 	    &              &   131.7543   &  131.5955	& 131.6060   &  131.6036   &   1.859-14    \\
  20 &  2s3d	  &  $^1$D$  _2$  & 	    &              &   132.4454   &  132.2961	& 132.3068   &  132.3025   &   2.393-14    \\
  21 &  2p3s	  &  $^3$P$^o_0$  & 	    &              &   132.7847   &  132.6789	& 132.7063   &  132.7065   &   1.657-13    \\
  22 &  2p3s	  &  $^3$P$^o_1$  & 	    &              &   133.0541   &  132.9407	& 132.9691   &  132.9681   &   1.259-13    \\
  23 &  2p3p	  &  $^3$D$  _1$  & 	    &              &   134.0904   &  133.9808	& 134.0095   &  134.0094   &   9.384-14    \\
  24 &  2p3p	  &  $^3$P$  _0$  & 	    &              &   135.2060   &  135.1068	& 135.1604   &  135.1568   &   5.901-14    \\
  25 &  2p3p	  &  $^1$P$  _1$  & 	    &              &   135.3264   &  135.2010	& 135.2351   &  135.2347   &   6.822-14    \\
  26 &  2p3p	  &  $^3$D$  _2$  & 	    &              &   135.3576   &  135.2211	& 135.2537   &  135.2528   &   7.587-14    \\
  27 &  2p3s	  &  $^3$P$^o_2$  & 	    &              &   135.6714   &  135.4818	& 135.5069   &  135.5070   &   9.257-14    \\
  28 &  2p3d	  &  $^3$F$^o_2$  & 	    &              &   135.9503   &  135.8111	& 135.8456   &  135.8439   &   1.429-13    \\
  29 &  2p3s	  &  $^1$P$^o_1$  & 	    &              &   136.1703   &  135.9858	& 136.0135   &  136.0108   &   1.083-13    \\
  30 &  2p3d	  &  $^3$F$^o_3$  & 	    &              &   136.6778   &  136.5256	& 136.5655   &  136.5632   &   3.471-14    \\
  31 &  2p3d	  &  $^3$D$^o_2$  & 	    &              &   136.7639   &  136.6154	& 136.6544   &  136.6544   &   2.380-14    \\
  32 &  2p3d	  &  $^3$D$^o_1$  & 	    &              &   136.8493   &  136.7010	& 136.7379   &  136.7377   &   1.543-14    \\
  33 &  2p3p	  &  $^3$P$  _1$  & 	    &              &   137.4442   &  137.2616	& 137.2978   &  137.2971   &   6.117-14    \\
  34 &  2p3p	  &  $^3$P$  _2$  & 	    &              &   137.6575   &  137.4645	& 137.5062   &  137.5043   &   5.651-14    \\
  35 &  2p3p	  &  $^3$D$  _3$  & 	    &              &   137.8754   &  137.6606	& 137.6824   &  137.6823   &   8.289-14    \\
  36 &  2p3p	  &  $^3$S$  _1$  & 	    &              &   137.9585   &  137.7517	& 137.7753   &  137.7751   &   6.455-14    \\
  37 &  2p3p	  &  $^1$D$  _2$  & 	    &              &   138.7900   &  138.5838	& 138.6296   &  138.6271   &   5.369-14    \\
  38 &  2p3d	  &  $^1$D$^o_2$  & 	    &              &   139.0858   &  138.8646	& 138.8996   &  138.8991   &   2.864-14    \\
  39 &  2p3d	  &  $^3$F$^o_4$  & 	    &              &   139.0920   &  138.8618	& 138.8920   &  138.8895   &   1.991-11    \\
  40 &  2p3d	  &  $^3$D$^o_3$  & 	    &              &   139.3807   &  139.1487	& 139.1842   &  139.1831   &   1.895-14    \\
\hline
\end{tabular} 
\end{table*}   

\clearpage  
\setcounter{table}{2} 
\begin{table*}   
\begin{tabular}{rllrrrrrrl} \hline
 Index  & Configuration       & Level & NIST & MBPT & GRASP1 & GRASP2     &  FAC1      & FAC2        & $\tau$ (s) \\ 
\hline
  41 &  2p3d	  &  $^3$P$^o_1$  & 	    &              &   139.6025   &  139.3823	& 139.4176   &  139.4176   &   1.996-14    \\
  42 &  2p3d	  &  $^3$P$^o_0$  & 	    &              &   139.6248   &  139.4278	& 139.4622   &  139.4620   &   2.437-14    \\
  43 &  2p3p	  &  $^1$S$  _0$  & 	    &              &   139.6518   &  139.4719	& 139.5251   &  139.5150   &   6.259-14    \\
  44 &  2p3d	  &  $^3$P$^o_2$  & 	    &              &   139.6640   &  139.4373	& 139.4706   &  139.4706   &   1.986-14    \\
  45 &  2p3d	  &  $^1$F$^o_3$  & 	    &              &   140.3344   &  140.0968	& 140.1331   &  140.1278   &   1.135-14    \\
  46 &  2p3d	  &  $^1$P$^o_1$  & 	    &              &   140.4891   &  140.2742	& 140.3078   &  140.3060   &   1.790-14    \\
  47 &  2s4s	  &  $^3$S$  _1$  & 	    &              &   171.5349   &  171.3948	& 171.3980   &  171.3969   &   2.004-13    \\
  48 &  2s4s	  &  $^1$S$  _0$  & 	    &              &   171.8124   &  171.6756	& 171.6823   &  171.6796   &   2.454-13    \\
  49 &  2s4p	  &  $^3$P$^o_0$  &   &   &  172.2952	  &  172.1604	 &  172.1693  &  172.1693   &  4.057-13    \\
  50 &  2s4p	  &  $^3$P$^o_1$  &   &   &  172.3205	  &  172.1817	 &  172.1910  &  172.1903   &  1.607-13    \\
  51 &  2s4p	  &  $^3$P$^o_2$  &   &   &  172.6433	  &  172.4966	 &  172.5040  &  172.5042   &  4.333-13    \\
  52 &  2s4p	  &  $^1$P$^o_1$  &   &   &  172.6955	  &  172.5474	 &  172.5566  &  172.5546   &  7.385-14    \\
  53 &  2s4d	  &  $^3$D$  _1$  &   &   &  173.0713	  &  172.9192	 &  172.9248  &  172.9236   &  4.396-14    \\
  54 &  2s4d	  &  $^3$D$  _2$  &   &   &  173.1060	  &  172.9508	 &  172.9563  &  172.9551   &  4.470-14    \\
  55 &  2s4d	  &  $^3$D$  _3$  &   &   &  173.1793	  &  173.0227	 &  173.0282  &  173.0270   &  4.573-14    \\
  56 &  2s4d	  &  $^1$D$  _2$  &   &   &  173.3942	  &  173.2410	 &  173.2448  &  173.2425   &  4.935-14    \\
  57 &  2s4f	  &  $^3$F$^o_2$  &   &   &  173.4504	  &  173.2973	 &  173.3016  &  173.2990   &  9.927-14    \\
  58 &  2s4f	  &  $^3$F$^o_3$  &   &   &  173.4647	  &  173.3084	 &  173.3128  &  173.3100   &  9.945-14    \\
  59 &  2s4f	  &  $^3$F$^o_4$  &   &   &  173.5014	  &  173.3446	 &  173.3487  &  173.3459   &  9.976-14    \\
  60 &  2s4f	  &  $^1$F$^o_3$  &   &   &  173.5498	  &  173.3952	 &  173.4003  &  173.3974   &  1.010-13    \\
  61 &  2p4s	  &  $^3$P$^o_0$  &   &   &  176.2662	  &  176.1341	 &  176.1631  &  176.1633   &  2.553-13    \\
  62 &  2p4s	  &  $^3$P$^o_1$  &   &   &  176.3284	  &  176.1939	 &  176.2237  &  176.2220   &  2.105-13    \\
  63 &  2p4p	  &  $^3$D$  _1$  &   &   &  176.8300	  &  176.6953	 &  176.7273  &  176.7274   &  1.074-13    \\
  64 &  2p4p	  &  $^3$P$  _0$  &   &   &  177.2203	  &  177.0916	 &  177.1357  &  177.1284   &  1.033-13    \\
  65 &  2p4p	  &  $^3$S$  _1$  &   &   &  177.3232	  &  177.1834	 &  177.2181  &  177.2166   &  1.048-13    \\
  66 &  2p4p	  &  $^3$D$  _2$  &   &   &  177.3510	  &  177.2064	 &  177.2416  &  177.2403   &  1.060-13    \\
  67 &  2p4d	  &  $^3$F$^o_2$  &   &   &  177.6001	  &  177.4533	 &  177.4840  &  177.4830   &  1.322-13    \\
  68 &  2p4d	  &  $^3$D$^o_2$  &   &   &  177.8708	  &  177.7217	 &  177.7517  &  177.7512   &  4.637-14    \\
  69 &  2p4d	  &  $^3$F$^o_3$  &   &   &  177.9047	  &  177.7533	 &  177.7831  &  177.7811   &  5.176-14    \\
  70 &  2p4d	  &  $^3$D$^o_1$  &   &   &  177.9089	  &  177.7607	 &  177.7895  &  177.7882   &  3.731-14    \\
  71 &  2p4f	  &  $^3$G$  _3$  &   &   &  178.0126	  &  177.8619	 &  177.8886  &  177.8860   &  1.001-13    \\
  72 &  2p4f	  &  $^3$D$  _2$  &   &   &  178.0515	  &  177.9017	 &  177.9289  &  177.9290   &  1.000-13    \\
  73 &  2p4f	  &  $^3$D$  _3$  &   &   &  178.0844	  &  177.9344	 &  177.9612  &  177.9614   &  1.001-13    \\
  74 &  2p4f	  &  $^3$G$  _4$  &   &   &  178.0891	  &  177.9380	 &  177.9650  &  177.9623   &  1.020-13    \\
  75 &  2p4s	  &  $^3$P$^o_2$  &   &   &  179.2174	  &  179.0020	 &  179.0261  &  179.0262   &  2.736-13    \\
  76 &  2p4s	  &  $^1$P$^o_1$  &   &   &  179.3263	  &  179.1121	 &  179.1379  &  179.1337   &  1.765-13    \\
  77 &  2p4p	  &  $^3$P$  _1$  &   &   &  179.9074	  &  179.6942	 &  179.7251  &  179.7240   &  9.880-14    \\
  78 &  2p4p	  &  $^1$D$  _2$  &   &   &  179.9606	  &  179.7436	 &  179.7764  &  179.7746   &  1.001-13    \\
  79 &  2p4p	  &  $^3$D$  _3$  &   &   &  180.1395	  &  179.9136	 &  179.9397  &  179.9399   &  1.132-13    \\
  80 &  2p4p	  &  $^1$P$  _1$  &   &   &  180.1659	  &  179.9429	 &  179.9701  &  179.9677   &  9.841-14    \\
\hline
\end{tabular} 
\end{table*}   

\clearpage  
\setcounter{table}{2} 
\begin{table*}   
\begin{tabular}{rllrrrrrrl} \hline
 Index  & Configuration       & Level & NIST & MBPT & GRASP1 & GRASP2     &  FAC1      & FAC2        & $\tau$ (s) \\ 
\hline

  81 &  2p4p	  &  $^3$P$  _2$  &   &   &  180.4185	  &  180.1963	 &  180.2311  &  180.2285   &  9.909-14    \\
  82 &  2p4d	  &  $^1$D$^o_2$  &   &   &  180.6077	  &  180.3794	 &  180.4051  &  180.4046   &  6.246-14    \\
  83 &  2p4d	  &  $^3$F$^o_4$  &   &   &  180.6447	  &  180.4125	 &  180.4386  &  180.4371   &  2.002-13    \\
  84 &  2p4d	  &  $^3$D$^o_3$  &   &   &  180.7003	  &  180.4682	 &  180.4935  &  180.4924   &  4.998-14    \\
  85 &  2p4p	  &  $^1$S$  _0$  &   &   &  180.7277	  &  180.5151	 &  180.5569  &  180.5381   &  1.145-13    \\
  86 &  2p4d	  &  $^3$P$^o_1$  &   &   &  180.7771	  &  180.5492	 &  180.5744  &  180.5732   &  4.456-14    \\
  87 &  2p4d	  &  $^3$P$^o_0$  &   &   &  180.7878	  &  180.5683	 &  180.5936  &  180.5917   &  5.243-14    \\
  88 &  2p4d	  &  $^3$P$^o_2$  &   &   &  180.8098	  &  180.5796	 &  180.6045  &  180.6036   &  4.627-14    \\
  89 &  2p4f	  &  $^3$F$  _3$  &   &   &  180.9192	  &  180.6877	 &  180.7099  &  180.7088   &  9.974-14    \\
  90 &  2p4f	  &  $^3$F$  _4$  &   &   &  180.9668	  &  180.7341	 &  180.7562  &  180.7545   &  1.004-13    \\
  91 &  2p4f	  &  $^3$F$  _2$  &   &   &  180.9906	  &  180.7605	 &  180.7832  &  180.7833   &  9.969-14    \\
  92 &  2p4f	  &  $^1$F$  _3$  &   &   &  181.0208	  &  180.7901	 &  180.8123  &  180.8124   &  9.990-14    \\
  93 &  2p4d	  &  $^1$F$^o_3$  &   &   &  181.0308	  &  180.7972	 &  180.8205  &  180.8160   &  2.891-14    \\
  94 &  2p4f	  &  $^3$G$  _5$  &   &   &  181.0437	  &  180.8091	 &  180.8310  &  180.8263   &  1.008-13    \\
  95 &  2p4f	  &  $^1$G$  _4$  &   &   &  181.0536	  &  180.8190	 &  180.8420  &  180.8374   &  1.033-13    \\
  96 &  2p4f	  &  $^3$D$  _1$  &   &   &  181.0726	  &  180.8457	 &  180.8682  &  180.8684   &  9.919-14    \\
  97 &  2p4d	  &  $^1$P$^o_1$  &   &   &  181.0995	  &  180.8745	 &  180.8978  &  180.8941   &  4.107-14    \\
  98 &  2p4f	  &  $^1$D$  _2$  &   &   &  181.1437	  &  180.9151	 &  180.9379  &  180.9379   &  1.002-13    \\
 \hline	
\end{tabular}

\begin{flushleft}
{\small
NIST: {\tt http://www.nist.gov/pml/data/asd.cfm} \\
MBPT: Gu \cite{gu} for the lowest 10 levels and Safronova {\em et al} \cite{uis4}  for the remaining levels \\
GRASP1: Coulomb energies \\
GRASP2: QED corrected energies \\
FAC1: Energies from the FAC for 98 level calculations\\
FAC2: Energies from the FAC for 166 level calculations \\
}
\end{flushleft}
                 
\end{table*}                                                                                                                                                    

\clearpage
\setcounter{table}{6} 
\begin{table*}             
\caption{Comparisons of oscillator strengths (f- values) for E1 transitions from the lowest 5 levels of Cl XIV and K XVI. $a{\pm}b \equiv a{\times}$10$^{{\pm}b}$. }
\begin{tabular}{rrlllllllll} \hline
\multicolumn{2}{c}{Transition} & \multicolumn{3}{c}{Cl XIV} & \multicolumn{3}{c}{K XVI}  \\
  \hline
   I    &   J  &   GRASP     &  MCHF       & R      &  GRASP      &  MCHF       &    R    \\
 \hline
    1   &   3  &  1.534$-$4  &  1.688$-$4  &  0.81  &  2.901$-$4  &  3.148$-$4  &   0.85  \\
    1   &   5  &  2.216$-$1  &  2.173$-$1  &  0.98  &  1.992$-$1  &  1.958$-$1  &   0.99  \\
    1   &  13  &  4.503$-$1  &  4.867$-$1  &  0.98  &  3.753$-$1  &  4.093$-$1  &   0.98  \\
    1   &  15  &  1.643$-$1  &  1.318$-$1  &  0.98  &  2.581$-$1  &  2.272$-$1  &   0.98  \\
    1   &  22  &  1.584$-$3  &  1.609$-$3  &  1.10  &  3.086$-$3  &  3.095$-$3  &   1.00  \\
    1   &  24  &  2.692$-$2  &  2.326$-$2  &  1.10  &  2.769$-$2  &  2.431$-$2  &   1.10  \\
\hline										                           
\end{tabular} 

\begin{flushleft}
{\small
GRASP: Present results from the {\sc grasp} code  \\
MCHF: Earlier results of C Froese Fischer (2009) from the {\sc mchf} code available at the \\website {\tt http://nlte.nist.gov/MCHF/view.html} \\
R: ratio of the velocity and length form of the f- values from the {\sc grasp} code \\
}
\end{flushleft}
\end{table*} 

\clearpage
\setcounter{table}{7} 
\begin{table*}             
\caption{Comparisons of radiative rates (A- values, s$^{-1}$) for M1 and M2 transitions  of Cl XIV and K XVI. $a{\pm}b \equiv a{\times}$10$^{{\pm}b}$.} 
\begin{tabular}{rrlllllllll}  \hline
\multicolumn{2}{c}{Transition} & Type & \multicolumn{2}{c}{Cl XIV} & \multicolumn{2}{c}{K XVI}  \\
 \hline
   I    &   J  &   & GRASP     &  MBPT        &  GRASP      &  MBPT     \\
\hline
  1  &  4  &  M2  &  5.96$-$1  &  5.85$-$1  &  1.05$-$0  &  1.03$-$0 \\
  1  &  7  &  M1  &  1.55$+$2  &  1.49$+$2  &  4.79$+$2  &  4.63$+$2 \\
  2  &  3  &  M1  &  3.13$-$0  &  3.16$-$0  &  1.32$+$1  &  1.32$+$1 \\
  2  &  5  &  M1  &  3.19$+$2  &  2.60$+$2  &  1.04$+$3  &  8.71$+$2 \\
  3  &  4  &  M1  &  2.83$+$1  &  2.89$+$1  &  1.33$+$2  &  1.36$+$2 \\
  3  &  5  &  M1  &  2.28$+$2  &  1.85$+$2  &  7.14$+$2  &  5.98$+$2 \\
  4  &  5  &  M1  &  3.10$+$2  &  2.50$+$2  &  8.92$+$2  &  7.37$+$2 \\
  6  &  7  &  M1  &  7.40$-$0  &  7.79$-$0  &  4.11$+$1  &  4.33$+$1 \\
  7  &  8  &  M1  &  1.51$+$1  &  1.50$+$1  &  5.72$+$1  &  5.60$+$1 \\
  7  &  9  &  M1  &  1.05$+$2  &  6.99$+$1  &  3.95$+$2  &  2.84$+$2 \\
  7  & 10  &  M1  &  3.24$+$3  &  2.55$+$3  &  9.97$+$3  &  8.15$+$3 \\
  8  &  9  &  M1  &  1.94$+$2  &  1.25$+$2  &  6.09$+$2  &  4.27$+$2 \\
 \hline										                           
\end{tabular} 

\begin{flushleft}
{\small
GRASP: Present results from the {\sc grasp} code  \\
MBPT: Results of Safronova {\em et al} \cite{uis3} from the {\em many-body perturbation theory} for M1 transitions and of  Safronova \cite{uis5} for the M2  \\
}
\end{flushleft}
\end{table*} 


\clearpage
\setcounter{table}{8} 
\begin{table*}             
\caption{Comparison of lifetimes ($\tau$) among the lowest 10 levels of Cl XIV. } 

\begin{tabular}{rrlllllllll} \hline
Level &  GRASP & MBPT & Experimental \\
\hline
   3  &  605.2 ns    &  ...  &  ...   \\
   4  &  34.64 ms    &  ...  &  ...    \\
   5  &  110.0 ps    &  116$^a$  & 150$\pm$30 \cite{jpf}, 140$\pm$10 \cite{kk}, 117$\pm$10 \cite{prl}    \\
   6  &  150.0 ps    &  227  & 170$\pm$9 \cite{prl}    \\
   7  &  144.9 ps    &  224  & 170$\pm$10 \cite{jpf}, 160$\pm$10 \cite{kk}, 165$\pm$15 \cite{prl}    \\
   8  &  141.5 ps    &  220  & 160$\pm$10 \cite{jpf}, 169$\pm$5 \cite{kk}, 160$\pm$10 \cite{prl}   \\
   9  &  567.7 ps    &  ...  & 600$\pm$60 \cite{jpf}, 630$\pm$20 \cite{kk} 625$\pm$25 \cite{prl}, 540$\pm$10 \cite{ishi}, 600 \cite{saka} \\
  10  &  70.51 ps    &  ...  & 84$\pm$5 \cite{jpf}, 77$\pm$9 \cite{prl}    \\
 \hline										                           
\end{tabular} 

\begin{flushleft}
{\small
GRASP: present results with the {\sc grasp} code \\
MBPT: Saforonova {\em et al} \cite{uis2} with the MBPT code \\
$a$:   Andersson {\em et al} \cite{azh} \\

}
\end{flushleft}
\end{table*}


\clearpage
\setcounter{table}{9}     
\begin{table*}      
\caption{Collision strengths for the resonance transitions of  Cl XIV. ($a{\pm}b \equiv$ $a\times$10$^{{\pm}b}$).}          
\begin{tabular}{rrllllllll}                                                                                   
\hline                                                                                                                                                                                                              
\multicolumn{2}{c}{Transition} & \multicolumn{7}{c}{Energy (Ryd)}\\                                           
\hline                                                                                                        
  $i$ & $j$ &    100 & 200 & 300 & 400 & 500 & 600 & 700   \\                                                 
\hline    
  1 &  2 &  7.813$-$04 &  3.038$-$04 &  1.605$-$04 &  9.920$-$05 &  6.739$-$05 &  4.870$-$05 &  3.698$-$05 \\
  1 &  3 &  4.405$-$03 &  2.874$-$03 &  2.416$-$03 &  2.235$-$03 &  2.018$-$03 &  1.917$-$03 &  1.851$-$03 \\
  1 &  4 &  3.859$-$03 &  1.497$-$03 &  7.903$-$04 &  4.880$-$04 &  3.314$-$04 &  2.394$-$04 &  1.817$-$04 \\
  1 &  5 &  1.202$-$00 &  1.282$-$00 &  1.281$-$00 &  1.332$-$00 &  1.221$-$00 &  1.188$-$00 &  1.176$-$00 \\
  1 &  6 &  4.014$-$05 &  2.490$-$05 &  2.097$-$05 &  1.940$-$05 &  1.862$-$05 &  1.819$-$05 &  1.790$-$05 \\
  1 &  7 &  5.330$-$05 &  1.470$-$05 &  6.014$-$06 &  3.037$-$06 &  1.750$-$06 &  1.100$-$06 &  7.350$-$07 \\
\hline                                                                                                        
\end{tabular}                                                                                                 
\end{table*}                                                                                                  
\newpage                                                                                                      
\clearpage                                                                                                    
\setcounter{table}{9}                                                                                         
\begin{table*}                                                                                                
\caption{Collision strengths for the resonance transitions of  Cl XIV. ($a{\pm}b \equiv$ $a\times$10$^{{\pm}b}$).}          
\begin{tabular}{rrllllllll}                                                                                   
\hline                                                                                                        
\multicolumn{2}{c}{Transition} & \multicolumn{7}{c}{Energy (Ryd)}\\                                           
\hline                                                                                                        
  $i$ & $j$ &    100 & 200 & 300 & 400 & 500 & 600 & 700   \\                                                 
\hline                                                                                                        
  1 & 47 &  2.386$-$04 &  7.346$-$05 &  3.558$-$05 &  2.099$-$05 &  1.389$-$05 &  9.847$-$06 &  7.396$-$06 \\
  1 & 48 &  7.479$-$03 &  8.263$-$03 &  8.565$-$03 &  8.737$-$03 &  8.860$-$03 &  8.957$-$03 &  9.040$-$03 \\
  1 & 49 &  7.504$-$05 &  1.935$-$05 &  8.501$-$06 &  4.711$-$06 &  2.983$-$06 &  2.044$-$06 &  1.486$-$06 \\
  1 & 50 &  7.339$-$04 &  9.332$-$04 &  1.144$-$03 &  1.318$-$03 &  1.461$-$03 &  1.577$-$03 &  1.685$-$03 \\
  1 & 51 &  3.720$-$04 &  9.588$-$05 &  4.210$-$05 &  2.332$-$05 &  1.476$-$05 &  1.011$-$05 &  7.352$-$06 \\
  1 & 52 &  1.080$-$02 &  1.838$-$02 &  2.351$-$02 &  2.742$-$02 &  3.057$-$02 &  3.308$-$02 &  3.543$-$02 \\
\hline                                                                                                        
\end{tabular}                                                                                                 
\end{table*}                                                                                                  
\newpage                                                                                                      
\clearpage                                                                                                    
\setcounter{table}{9}                                                                                         
\begin{table*}                                                                                                
\caption{Collision strengths for the resonance transitions of  Cl XIV. ($a{\pm}b \equiv$ $a\times$10$^{{\pm}b}$).}          
\begin{tabular}{rrllllllll}                                                                                   
\hline                                                                                                        
\multicolumn{2}{c}{Transition} & \multicolumn{7}{c}{Energy (Ryd)}\\                                           
\hline                                                                                                        
  $i$ & $j$ &    100 & 200 & 300 & 400 & 500 & 600 & 700   \\                                                 
\hline                                                                                                        
  1 & 92 &  4.113$-$06 &  1.078$-$06 &  5.710$-$07 &  3.861$-$07 &  2.925$-$07 &  2.360$-$07 &  1.983$-$07 \\
  1 & 93 &  2.183$-$05 &  2.912$-$05 &  3.311$-$05 &  3.552$-$05 &  3.700$-$05 &  3.802$-$05 &  3.883$-$05 \\
  1 & 94 &  1.830$-$05 &  2.121$-$05 &  2.312$-$05 &  2.435$-$05 &  2.520$-$05 &  2.584$-$05 &  2.635$-$05 \\
  1 & 95 &  2.100$-$06 &  4.082$-$07 &  1.587$-$07 &  8.248$-$08 &  5.014$-$08 &  3.360$-$08 &  2.405$-$08 \\
  1 & 96 &  5.752$-$05 &  8.130$-$05 &  9.300$-$05 &  9.990$-$05 &  1.041$-$04 &  1.070$-$04 &  1.093$-$04 \\
  1 & 97 &  6.336$-$05 &  6.949$-$05 &  7.357$-$05 &  7.649$-$05 &  7.871$-$05 &  8.053$-$05 &  8.207$-$05 \\
  1 & 98 &  9.632$-$04 &  1.357$-$03 &  1.615$-$03 &  1.811$-$03 &  1.971$-$03 &  2.103$-$03 &  2.226$-$03 \\
\hline                                                                                                        
\end{tabular}                                                                                                 
\end{table*}      
                                                                                                                                                                                            

\newpage                                                                                                      
\clearpage                                                                                                    
\setcounter{table}{10}                                                                                         
\begin{table*}                                                                                                
\caption{Collision strengths for the resonance transitions of  K XVI. ($a{\pm}b \equiv$ $a\times$10$^{{\pm}b}$).}                
\begin{tabular}{rrlllllllll}                                                                                   
\hline                                                                                                                                                                                                               
\multicolumn{2}{c}{Transition} & \multicolumn{8}{c}{Energy (Ryd)}\\                                           
\hline                                                                                                        
  $i$ & $j$ &    100 & 200 & 300 & 400 & 500 & 600 & 700  & 800  \\                                                 
\hline    
  1 &  2 &  7.875$-$04 &  3.346$-$04 &  1.837$-$04 &  1.160$-$04 &  7.985$-$05 &  5.847$-$05 &  4.455$-$05 &  3.512$-$05 \\
  1 &  3 &  5.760$-$03 &  4.449$-$03 &  3.784$-$03 &  3.568$-$03 &  3.410$-$03 &  3.177$-$03 &  3.142$-$03 &  3.018$-$03 \\
  1 &  4 &  3.874$-$03 &  1.641$-$03 &  8.995$-$04 &  5.674$-$04 &  3.904$-$04 &  2.858$-$04 &  2.177$-$04 &  1.716$-$04 \\
  1 &  5 &  9.055$-$01 &  1.050$-$00 &  1.013$-$00 &  1.046$-$00 &  1.034$-$00 &  9.672$-$01 &  9.974$-$01 &  9.398$-$01 \\
  1 &  6 &  5.350$-$05 &  3.677$-$05 &  3.175$-$05 &  2.960$-$05 &  2.849$-$05 &  2.782$-$05 &  2.741$-$05 &  2.718$-$05 \\
  1 &  7 &  5.856$-$05 &  1.843$-$05 &  8.047$-$06 &  4.234$-$06 &  2.502$-$06 &  1.603$-$06 &  1.091$-$06 &  7.807$-$07 \\
\hline                                                                                                        
\end{tabular}                                                                                                 
\end{table*}                                                                                                  
\newpage                                                                                                      
\clearpage                                                                                                    
\setcounter{table}{10}                                                                                        
\begin{table*}                                                                                                
\caption{Collision strengths for the resonance transitions of  K XVI. ($a{\pm}b \equiv$ $a\times$10$^{{\pm}b}$).}           
\begin{tabular}{rrlllllllll}                                                                                  
\hline                                                                                                        
\multicolumn{2}{c}{Transition} & \multicolumn{8}{c}{Energy (Ryd)}\\                                           
\hline                                                                                                        
  $i$ & $j$ &    100 & 200 & 300 & 400 & 500 & 600 & 700  & 800    \\                                         
\hline                                                                                                        
  1 & 47 &  2.794$-$04 &  8.932$-$05 &  4.404$-$05 &  2.622$-$05 &  1.739$-$05 &  1.248$-$05 &  9.321$-$06 &  7.285$-$06 \\
  1 & 48 &  5.580$-$03 &  6.304$-$03 &  6.594$-$03 &  6.755$-$03 &  6.864$-$03 &  6.950$-$03 &  7.020$-$03 &  7.084$-$03 \\
  1 & 49 &  9.241$-$05 &  2.516$-$05 &  1.120$-$05 &  6.244$-$06 &  3.946$-$06 &  2.716$-$06 &  1.973$-$06 &  1.502$-$06 \\
  1 & 50 &  8.120$-$04 &  1.073$-$03 &  1.343$-$03 &  1.566$-$03 &  1.752$-$03 &  1.911$-$03 &  2.061$-$03 &  2.162$-$03 \\
  1 & 51 &  4.569$-$04 &  1.243$-$04 &  5.529$-$05 &  3.082$-$05 &  1.947$-$05 &  1.340$-$05 &  9.730$-$06 &  7.403$-$06 \\
  1 & 52 &  6.654$-$03 &  1.209$-$02 &  1.586$-$02 &  1.875$-$02 &  2.111$-$02 &  2.310$-$02 &  2.497$-$02 &  2.622$-$02 \\
\hline                                                                                                        
\end{tabular}                                                                                                 
\end{table*}                                                                                                  
\newpage                                                                                                      
\clearpage                                                                                                    
\setcounter{table}{10}                                                                                        
\begin{table*}                                                                                                
\caption{Collision strengths for the resonance transitions of  K XVI. ($a{\pm}b \equiv$ $a\times$10$^{{\pm}b}$).}           
\begin{tabular}{rrlllllllll}                                                                                  
\hline                                                                                                        
\multicolumn{2}{c}{Transition} & \multicolumn{8}{c}{Energy (Ryd)}\\                                           
\hline                                                                                                        
  $i$ & $j$ &    100 & 200 & 300 & 400 & 500 & 600 & 700  & 800    \\                                         
\hline                                                                                                        
  1 & 92 &  5.094$-$06 &  1.081$-$06 &  4.724$-$07 &  2.672$-$07 &  1.723$-$07 &  1.204$-$07 &  8.886$-$08 &  6.824$-$08 \\
  1 & 93 &  1.562$-$05 &  2.113$-$05 &  2.466$-$05 &  2.688$-$05 &  2.838$-$05 &  2.943$-$05 &  3.032$-$05 &  3.086$-$05 \\
  1 & 94 &  1.272$-$05 &  1.441$-$05 &  1.590$-$05 &  1.692$-$05 &  1.765$-$05 &  1.821$-$05 &  1.865$-$05 &  1.901$-$05 \\
  1 & 95 &  3.013$-$06 &  6.025$-$07 &  2.343$-$07 &  1.211$-$07 &  7.326$-$08 &  4.886$-$08 &  3.485$-$08 &  2.607$-$08 \\
  1 & 96 &  3.855$-$05 &  5.779$-$05 &  6.815$-$05 &  7.447$-$05 &  7.872$-$05 &  8.164$-$05 &  8.412$-$05 &  8.558$-$05 \\
  1 & 97 &  3.947$-$05 &  4.424$-$05 &  4.774$-$05 &  5.020$-$05 &  5.207$-$05 &  5.356$-$05 &  5.482$-$05 &  5.589$-$05 \\
  1 & 98 &  6.016$-$04 &  8.693$-$04 &  1.044$-$03 &  1.177$-$03 &  1.285$-$03 &  1.378$-$03 &  1.460$-$03 &  1.525$-$03 \\
\hline                                                                                                        
\end{tabular}                                                                                                 
\end{table*}                                                                                                  


\newpage                                                                                                      
\clearpage                                                                                                    
\setcounter{table}{11}                                                                                        
\begin{table*}                                                                                                
\caption{Collision strengths for the resonance transitions of  Ge XXIX. ($a{\pm}b \equiv$ $a\times$10$^{{\pm}b}$).}         
\begin{tabular}{rrlllllllll}                                                                                  
\hline                                                                                                        
\multicolumn{2}{c}{Transition} & \multicolumn{8}{c}{Energy (Ryd)}\\                                           
\hline                                                                                                        
  $i$ & $j$ &    500 & 800 & 1100 & 1400 & 1700 & 2000 & 2300 & 2600    \\                                    
\hline    
  1 &  2 &  1.385$-$04 &  7.256$-$05 &  4.465$-$05 &  3.024$-$05 &  2.186$-$05 &  1.656$-$05 &  1.294$-$05 &  1.041$-$05 \\
  1 &  3 &  2.238$-$02 &  1.918$-$02 &  1.946$-$02 &  1.889$-$02 &  1.928$-$02 &  1.954$-$02 &  1.912$-$02 &  1.803$-$02 \\
  1 &  4 &  6.449$-$04 &  3.358$-$04 &  2.059$-$04 &  1.390$-$04 &  1.003$-$04 &  7.589$-$05 &  5.924$-$05 &  4.761$-$05 \\
  1 &  5 &  3.741$-$01 &  3.215$-$01 &  3.278$-$01 &  3.188$-$01 &  3.408$-$01 &  3.478$-$01 &  3.459$-$01 &  3.045$-$01 \\
  1 &  6 &  9.350$-$05 &  8.912$-$05 &  8.749$-$05 &  8.697$-$05 &  8.698$-$05 &  8.743$-$05 &  8.801$-$05 &  8.883$-$05 \\
  1 &  7 &  8.877$-$06 &  3.896$-$06 &  2.093$-$06 &  1.278$-$06 &  8.490$-$07 &  5.981$-$07 &  4.425$-$07 &  3.389$-$07 \\
\hline                                                                                                        
\end{tabular}                                                                                                 
\end{table*}                                                                                                  
\newpage                                                                                                      
\clearpage                                                                                                    
\setcounter{table}{11}                                                                                        
\begin{table*}                                                                                                
\caption{Collision strengths for the resonance transitions of  Ge XXIX. ($a{\pm}b \equiv$ $a\times$10$^{{\pm}b}$).}         
\begin{tabular}{rrlllllllll}                                                                                  
\hline                                                                                                        
\multicolumn{2}{c}{Transition} & \multicolumn{8}{c}{Energy (Ryd)}\\                                           
\hline                                                                                                        
  $i$ & $j$ &    500 & 800 & 1100 & 1400 & 1700 & 2000 & 2300 & 2600    \\                                    
\hline                                                                                                        
  1 & 47 &  4.423$-$05 &  1.984$-$05 &  1.134$-$05 &  7.334$-$06 &  5.165$-$06 &  3.844$-$06 &  2.954$-$06 &  2.361$-$06 \\
  1 & 48 &  2.056$-$03 &  2.198$-$03 &  2.281$-$03 &  2.344$-$03 &  2.397$-$03 &  2.446$-$03 &  2.491$-$03 &  2.535$-$03 \\
  1 & 49 &  1.332$-$05 &  5.396$-$06 &  2.852$-$06 &  1.748$-$06 &  1.178$-$06 &  8.400$-$07 &  6.316$-$07 &  4.899$-$07 \\
  1 & 50 &  1.038$-$03 &  1.429$-$03 &  1.733$-$03 &  1.984$-$03 &  2.204$-$03 &  2.377$-$03 &  2.585$-$03 &  2.689$-$03 \\
  1 & 51 &  6.420$-$05 &  2.593$-$05 &  1.367$-$05 &  8.363$-$06 &  5.625$-$06 &  4.006$-$06 &  3.007$-$06 &  2.330$-$06 \\
  1 & 52 &  2.911$-$03 &  4.117$-$03 &  5.036$-$03 &  5.792$-$03 &  6.450$-$03 &  6.967$-$03 &  7.587$-$03 &  7.899$-$03 \\
\hline                                                                                                        
\end{tabular}                                                                                                 
\end{table*}                                                                                                  
\newpage                                                                                                      
\clearpage                                                                                                    
\setcounter{table}{11}                                                                                        
\begin{table*}                                                                                                
\caption{Collision strengths for the resonance transitions of  Ge XXIX. ($a{\pm}b \equiv$ $a\times$10$^{{\pm}b}$).}         
\begin{tabular}{rrlllllllll}                                                                                  
\hline                                                                                                        
\multicolumn{2}{c}{Transition} & \multicolumn{8}{c}{Energy (Ryd)}\\                                           
\hline                                                                                                        
  $i$ & $j$ &    500 & 800 & 1100 & 1400 & 1700 & 2000 & 2300 & 2600    \\                                    
\hline                                                                                                        
  1 & 92 &  3.515$-$07 &  1.520$-$07 &  9.455$-$08 &  6.922$-$08 &  5.526$-$08 &  4.645$-$08 &  4.040$-$08 &  3.597$-$08 \\
  1 & 93 &  4.735$-$06 &  5.445$-$06 &  5.960$-$06 &  6.355$-$06 &  6.678$-$06 &  6.957$-$06 &  7.203$-$06 &  7.429$-$06 \\
  1 & 94 &  3.106$-$07 &  1.184$-$07 &  6.304$-$08 &  3.930$-$08 &  2.688$-$08 &  1.955$-$08 &  1.485$-$08 &  1.166$-$08 \\
  1 & 95 &  2.315$-$06 &  2.661$-$06 &  2.907$-$06 &  3.092$-$06 &  3.241$-$06 &  3.364$-$06 &  3.474$-$06 &  3.571$-$06 \\
  1 & 96 &  2.216$-$07 &  7.625$-$08 &  3.718$-$08 &  2.173$-$08 &  1.418$-$08 &  9.959$-$09 &  7.370$-$09 &  5.668$-$09 \\
  1 & 97 &  1.143$-$04 &  1.430$-$04 &  1.647$-$04 &  1.825$-$04 &  1.979$-$04 &  2.107$-$04 &  2.248$-$04 &  2.343$-$04 \\
  1 & 98 &  1.087$-$05 &  1.372$-$05 &  1.547$-$05 &  1.668$-$05 &  1.767$-$05 &  1.834$-$05 &  1.904$-$05 &  1.943$-$05 \\
\hline                                                                                                        
\end{tabular}                                                                                                 
\end{table*}                                                                                                  


\clearpage      
\newpage                
\setcounter{table}{15}    
\begin{table*}   
\caption{Comparisons of effective collision strengths for transitions among the lowest 10 levels of Cl XIV. $a{\pm}b \equiv a{\times}$10$^{{\pm}b}$.} 
\begin{tabular}{rrlllllllll} \hline
\multicolumn{2}{c}{Transition} & \multicolumn{3}{c}{RM (log T$_e$, K)} & \multicolumn{3}{c}{DARC (log T$_e$, K)} & \multicolumn{3}{c}{Ratio R (log T$_e$, K)} \\
 \hline
   I    &   J  &      6.30     &     6.50    &     6.70     &   6.30      &     6.50    &    6.70     &   6.30 &  6.50  &  6.70 \\
 \hline
    1   &   2  &   3.605$-$3   &  3.104$-$3  &   2.610$-$3  &  5.797$-$3  &  5.271$-$3  &  4.459$-$3  &  1.61  &  1.70  & 1.71  \\
    1   &   3  &   1.237$-$2   &  1.096$-$2  &   9.588$-$3  &  1.901$-$2  &  1.750$-$2  &  1.511$-$2  &  1.54  &  1.60  & 1.58  \\
    1   &   4  &   1.815$-$2   &  1.556$-$2  &   1.304$-$2  &  2.978$-$2  &  2.711$-$2  &  2.292$-$2  &  1.64  &  1.74  & 1.76  \\
    1   &   5  &   8.039$-$1   &  8.567$-$1  &   9.185$-$1  &  7.984$-$1  &  8.391$-$1  &  8.922$-$1  &  0.99  &  0.98  & 0.97  \\
    1   &   6  &   1.292$-$4   &  1.134$-$4  &   9.763$-$5  &  4.055$-$4  &  4.057$-$4  &  3.582$-$4  &  3.14  &  3.58  & 3.67  \\
    1   &   7  &   3.152$-$4   &  2.715$-$4  &   2.273$-$4  &  9.447$-$4  &  9.212$-$4  &  7.970$-$4  &  3.00  &  3.39  & 3.51  \\
\hline										                           
\end{tabular} 

\begin{flushleft}
{\small
RM: Earlier interpolated results of Keenan \cite{fpk1} \\
DARC: Present results from the DARC code  \\
Ratio R: DARC/RM \\
}
\end{flushleft}
\end{table*}      


\clearpage      
\newpage                
\setcounter{table}{16}    
\begin{table*}   
\caption{Comparisons of effective collision strengths for transitions among the lowest 10 levels of K XVI. $a{\pm}b \equiv a{\times}$10$^{{\pm}b}$. }
\begin{tabular}{rrlllllllll} \hline
\multicolumn{2}{c}{Transition} & \multicolumn{3}{c}{RM (log T$_e$, K)} & \multicolumn{3}{c}{DARC (log T$_e$, K)} & \multicolumn{3}{c}{Ratio R (log T$_e$, K)} \\
\hline
   I    &   J  &      6.50     &     6.70    &     6.90     &   6.50      &     6.70    &    6.90     &   6.50 &  6.70  &  6.90 \\
\hline
    1  &  2  & 2.475$-$3 &  2.134$-$3 &  1.788$-$3 &   4.566$-$3 & 3.973$-$3 & 3.241$-$3 & 1.84 &  1.86 & 1.81  \\
    1  &  3  & 9.706$-$3 &  8.872$-$3 &  8.079$-$3 &   1.625$-$2 & 1.458$-$2 & 1.251$-$2 & 1.67 &  1.64 & 1.55  \\
    1  &  4  & 1.253$-$2 &  1.072$-$2 &  8.919$-$3 &   2.484$-$2 & 2.132$-$2 & 1.722$-$2 & 1.98 &  1.99 & 1.93  \\
    1  &  5  & 6.596$-$1 &  7.096$-$1 &  7.664$-$1 &   6.621$-$1 & 7.043$-$1 & 7.614$-$1 & 1.00 &  0.99 & 0.99  \\
    1  &  6  & 1.178$-$4 &  1.034$-$4 &  8.983$-$5 &   3.380$-$4 & 3.166$-$4 & 2.684$-$4 & 2.87 &  3.06 & 2.99  \\
    1  &  7  & 2.362$-$4 &  1.996$-$4 &  1.636$-$4 &   7.496$-$4 & 6.867$-$4 & 5.662$-$4 & 3.17 &  3.44 & 3.46  \\
\hline										                           
\end{tabular} 

\begin{flushleft}
{\small
RM: Earlier interpolated results of Keenan \cite{fpk1} \\
DARC: Present results from the DARC code  \\
Ratio R: DARC/RM \\
}
\end{flushleft}
\end{table*}  

\end{document}